\documentclass[aps,prb,twocolumn,showpacs,floatfix,superscriptaddress]{revtex4-1}
\usepackage{amsmath,amssymb}
\usepackage{graphicx}
\usepackage{times}
\usepackage{note,a4mine}
\usepackage{epsfig,subfigure,overpic}
\usepackage{mathrsfs}
\usepackage{url}

\def \beq {\begin{equation}}
\def \edq {\end{equation}}
\def \bes {\begin{subequations}}
\def \eds {\end{subequations}}
\def \beqn {\begin{equation*}}
\def \edqn {\end{equation*}}

\def \dag {\dagger}

\def \up {\uparrow}
\def \down {\downarrow}

\def \sm {\sigma}

\def \bk {\bar{k}}

\def \bn {\bar{n}}

\def \bs {\bar{s}}
\def \btau {\bar{\tau}}
\def \balpha {\bar{\alpha}}

\def \veps {\varepsilon}
\def \weps {\widetilde{\varepsilon}}

\def \calh {{\cal{H}}}

\def \calg {{\cal{G}}}

\def \calf {{\cal{F}}}

\def \scrf {{\mathscr{F}}}

\def \hatc {\hat{c}}
\def \hatg {\hat{\text{g}}}
\def \hatf {\hat{f}}

\def \hatV {\hat{V}}
\def \hsm {\hat{\sigma}}

\def \wV {\widetilde{V}}

\def \wg {\widetilde{\text{g}}}
\def \wGamma {\widetilde{\Gamma}}

\def \hatsig {\widehat{\Sigma}}
\def \wsig {\widetilde{\Sigma}}
\def \brsig {\breve{\Sigma}}

\def \tr {\text{Tr}}

\def \wbb {\widetilde{b}}

\providecommand{\nbraket}[1]{\left\langle#1\right\rangle}
\providecommand{\dbraket}[1]{\langle\langle#1\rangle\rangle}

\begin{document}
\title{Josephson current in carbon nanotubes with spin-orbit interaction}
\author{Jong Soo Lim}
\affiliation{Departament de F\'{i}sica, Universitat de les Illes Balears,
  E-07122 Palma de Mallorca, Spain}
\author{Rosa L\'opez}
\affiliation{Departament de F\'{i}sica, Universitat de les Illes Balears,
  E-07122 Palma de Mallorca, Spain}
\affiliation{Institut de F\'{i}sica Interdisciplinar i de Sistemes Complexos
  IFISC (CSIC-UIB), E-07122 Palma de Mallorca, Spain}
\author{Mahn-Soo Choi}
\affiliation{Department of Physics, Korea University, Seoul 136-701, Korea}
\author{Ram\' on Aguado}
\affiliation{Instituto de Ciencia de Materiales de Madrid (ICMM-CSIC), Cantoblanco,
  28049 Madrid, Spain}
  \begin{abstract}
We demonstrate that curvature-induced spin-orbit (SO) coupling induces a $0-\pi$ transition in the Josephson current through a carbon nanotube quantum dot coupled to superconducting leads. In the non-interacting regime, the transition can be tuned by applying parallel magnetic field near the critical field where orbital states become degenerate. Moreover, the interplay between charging and SO effects in the Coulomb Blockade and cotunneling regimes leads to a rich phase diagram with well-defined (analytical) boundaries in parameter space. Finally, the $0$ phase always prevails in the Kondo regime. Our calculations are relevant in view of recent experimental advances in transport through ultra-clean carbon nanotubes.

\end{abstract}
\maketitle
The spectrum of quantum dots (QDs) defined in carbon nanotubes (NTs) is four-fold degenerate owing to spin and valley symmetry. Recently, Kuemmeth {\it et al} \cite{Kuemmeth:08} have demonstrated that the spin and valley degrees of freedom are coupled in NTs. This spin-orbit (SO) coupling breaks the four-fold degeneracy into two Kramers doublets (time-reversed electrons pairs).
From a different perspective, NTs are interesting because they can support supercurrents when coupled to superconductors~\cite{Kasumov:99,Morpurgo:99,Jarillo-Herrero:06,Cleziou:06}. These supercurrents mainly result from resonant transmission through discrete states confined to the QD, the so-called Andreev bound states (ABS) corresponding to entangled time-reversed electron-hole Kramers pairs \footnote{For a review, see S. De Franceschi, L. P. Kouwenhoven, C. Schonenberger and W. Wersndorfer, Nature Nanotech., {\bf 5}, 703 (2010).}. As both phenomena, SO and ABS, are related to time-reversed Kramers pairs, it is thus interesting to raise the following question: how are the ABS, and therefore the Josephson effect, affected by SO coupling in NTs? 
Here we address this question. Using various theoretical approaches we analyse this problem in all relevant transport regimes and demonstrate that the SO coupling is able to reverse the supercurrent, namely to induce a $0$ to $\pi$ transition, even in the non-interacting regime.
\begin{figure}
  \centering
  \includegraphics[width=0.45\textwidth]{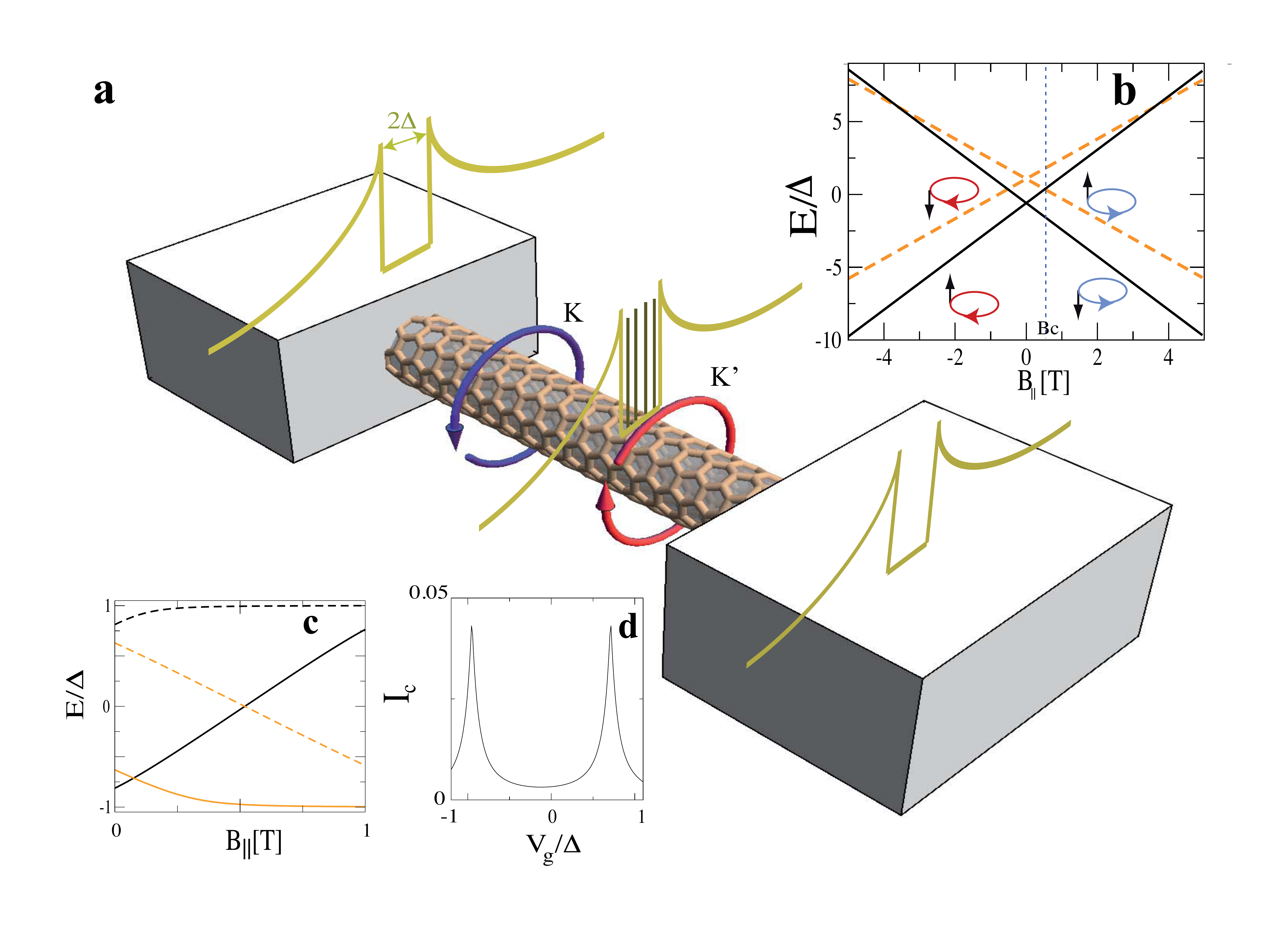}
  \caption{(Color online). (a) Schematics of a NT coupled to superconducting reservoirs. In the QD region, discrete Andreev levels form inside the BCS gap. The figure also show the K, K' orbits encircling the NT. (b) Energy spectrum of a NT QD for realistic experimental parameters (see supplementary info). All energies are given in units of the BCS gap $\Delta=0.25\text{meV}$, such that $\Delta_{SO}/\Delta \approx 1.66$) and referred to $E_F$ which we take as the energy at which ($K,\up$) and ($K',\up$) cross at $B_c\approx 0.52 T$ (dashed vertical line). (c) Andreev bound states corresponding to the spectrum in (b). Black (orange) lines correspond to ABS calculated from the lowest (highest) Kramers doublet (each contributes with two, solid and dashed, ABS). (d) Critical current (units $2e\Delta/\hbar$) versus gate voltage. The two peaks correspond to resonant Cooper pair tunneling through SO split Kramers pairs.}
  \label{fig:1}
\end{figure}

The valley isospin ($\tau=\pm$) originates from the two equivalent dispersion cones (K and K') in graphene, arising from time-inversion symmetry. When graphene is wrapped into a cylinder to create a NT, the valley degeneracy leads to two degenerate clockwise and counterclockwise electron orbits which encircle the NT. This degeneracy, together with spin, manifests in a four-fold shell structure in the Coulomb Blockade regime \cite{Liang:02,Cobden:02}, as well as in a SU(4) Kondo effect in the strongly correlated regime \cite{Jarillo-Herrero:05a,Choi:05}.  Furthermore, magnetic moments associated with these orbital persistent currents are remarkably large \cite{Minot:04} which allows to perform detailed transport spectroscopy when an external magnetic field is applied parallel to the NT axis \cite{Minot:04, Jarillo-Herrero:05a,Jarillo-Herrero:05b}. The orbital motion of electrons also 
couples to a curvature-induced radial electric field. This creates an effective axial magnetic field $B_{SO}$ which polarizes the spins along the NT axis and favors parallel alignment of the spin and orbital magnetic momenta ($K,\up$) and ($K',\down$) or antiparallel ($K,\down$) and ($K',\up$) depending on the sign of $\Delta_{SO}$.  As a result, the fourfold degeneracy breaks into two Kramers doublets (time-reversed electrons pairs) separated by an energy $\Delta_{SO}$  \footnote{Various band-structure calculations have been devoted to improve the first calculation in T. Ando, J. Phys. Soc. Jpn. {\bf 69}, 1757 (2000). See e.g. D. Huertas-hernando {\it et al}, Phys. Rev. B, {\bf 74}, 155426 (2006); L. Chico {\it et al}, Phys. Rev. B, {\bf 79} , 235423 (2009)}. Recent experiments \cite{Jespersen:11} have shown this SO effect also appears in disordered NTs in the multielectron regime. 
\begin{figure*}
\centering
\includegraphics[width=0.8\textwidth]{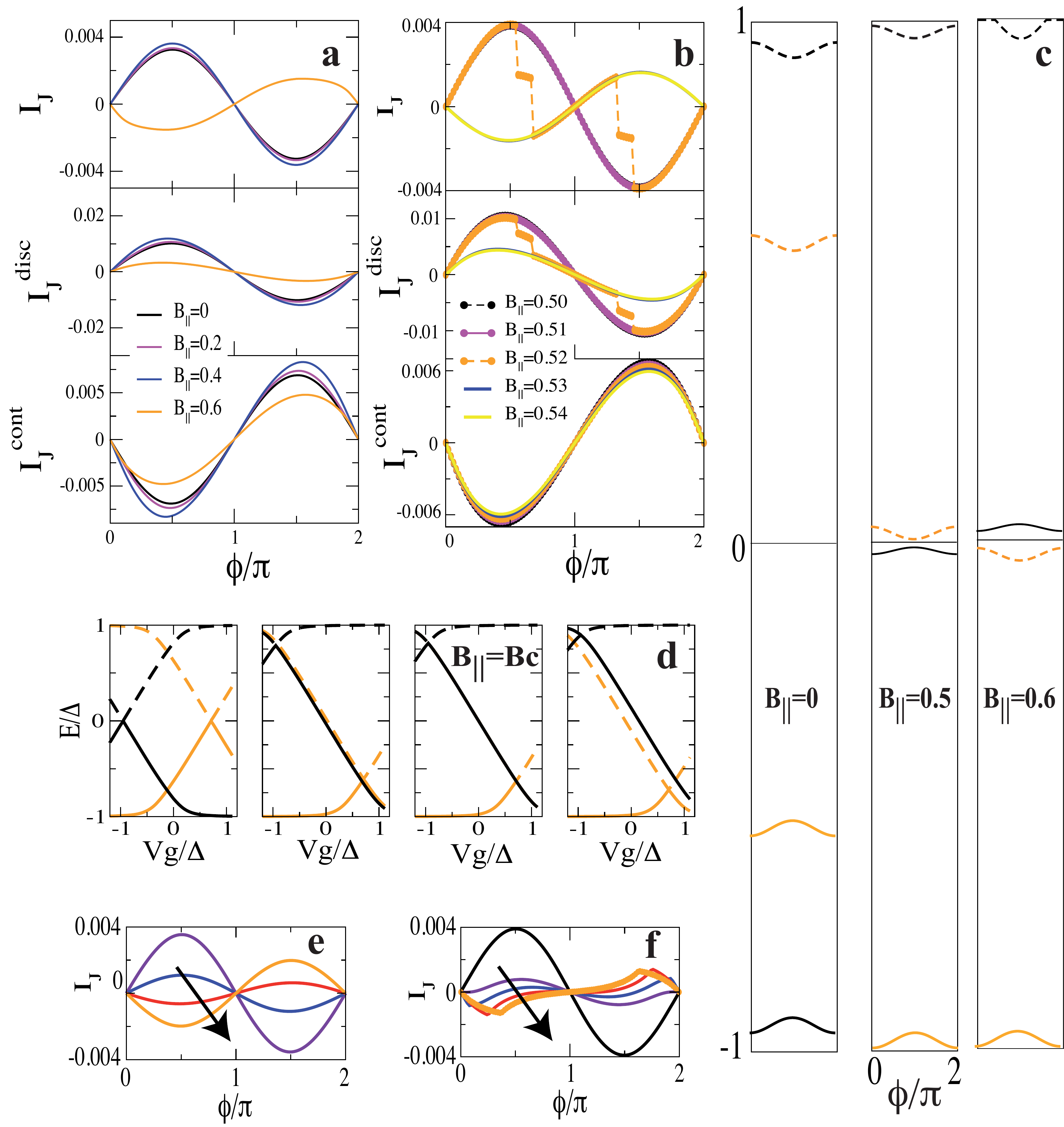}
\caption{(Color online) Total (top), discrete (middle) and continuous (bottom) Josephson current (units $2e\Delta/\hbar$) as a function of phase and different $B_{||}$ (in Tesla) for $\Gamma=0.1\Delta$. At the highest magnetic filed the system has $\pi$-junction behavior. (b) Same as (a) near the $0$-$\pi$ transition at $B_{||}=B_c=0.52T$.
(c) ABS vs. $\phi$ for different $B_{||}$. When $B_{||}\gtrsim B_c$, the two inner ABS cross at $E_F=0$ resulting in $\pi$ behavior.
(d) ABS versus $V_g$ for different $B_{||}=0,0.5,0.52$ and $0.6$ T, from left to right. At $B_{||}=B_c$ the two inner ABS are degenerate for all $|V_g|<\Delta$. The $\pi$ transition is robust as $V_g$ is varied (direction of the arrow) either above (e) or below (f) $V_g=0$.}
\label{fig:2}
\end{figure*}

The system we have in mind is shown in Fig. 1a. A QD NT with SO coupling is connected to superconducting leads with BCS density of states. Owing to the superconducting pairing, electrons in the NT with energies below the superconducting gap ($\Delta$) are reflected as their time-reversed particle, a hole with opposite spin and momentum. This process, known as Andreev reflection, leads to discrete states inside the gap, namely the ABS corresponding to entangled time-reversed electron-hole Kramers pairs. We model this system by an Anderson hamiltonian with s- wave superconducting reservoirs and with QD levels obtained from a NT model including SO. Green's function in Nambu representation are used to obtain the ABS and the two contributions to the Josephson current $I_J= I_J^{dis} + I_J^{con}$ of this model (full details are given in the supplementary info). The discrete part $I_J^{dis}$ is due to Cooper pair tunneling through the ABS and can be written as $I_J^{dis} = \frac{2e}{\hbar} \sum_{E_{1(2)}} f(E_{1(2)}) \frac{\partial E_{1(2)}}{\partial \phi}$, with $f(E)$ the Fermi-Dirac function. Namely, the derivative with respect to phase of the {\it occupied} ABS. The continuous part $I_J^{con}$ is due to particle-hole excitations for energies larger than $\Delta$. In the noninteracting case, the ABS can be obtained from
\begin{widetext}
\beq
\left(E_{1(2)} - \veps_{\mp\up} + \frac{\Gamma E_{1(2)}}{\sqrt{\Delta^2-{E^2_{1(2)}}}}\right)
\left(E_{1(2)} + \veps_{\pm\down} + \frac{\Gamma E_{1(2)}}{\sqrt{\Delta^2-{E^2_{1(2)}}}}\right)
- \frac{\Gamma^2\Delta^2\cos^2(\phi/2)}{\Delta^2-{E^2_{1(2)}}} 
= 0,
\label{eq:andreevbound}
\edq
\end{widetext}
where $\phi$ is the phase difference between superconductors and $\Gamma$ is the tunneling rate.
\begin{figure*}
\centering
\includegraphics[width=0.75\textwidth]{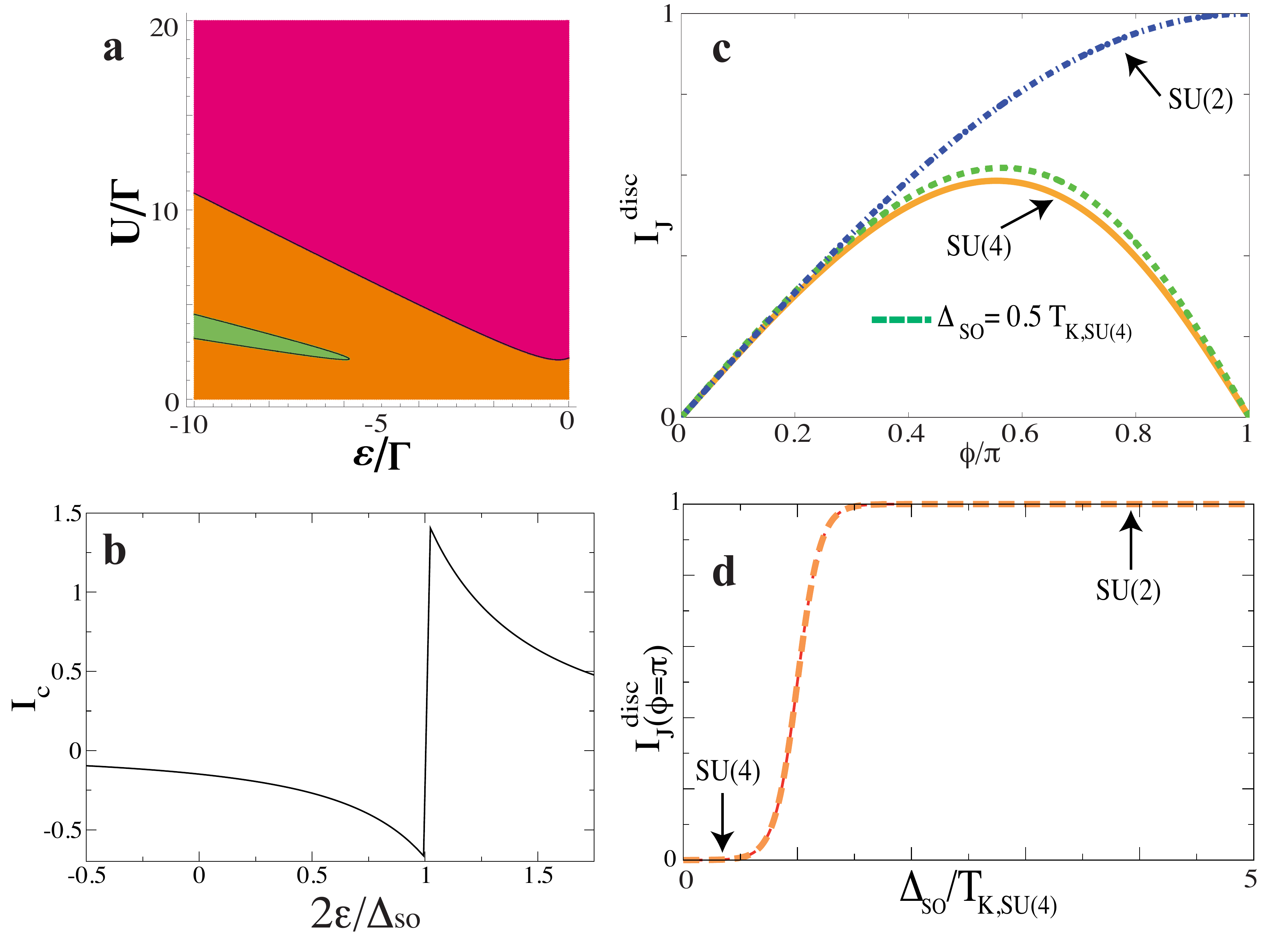}
\caption{(Color online). a) Phase diagram in the large gap limit and $\phi=0$. For large enough U, the ground state is $(S_z,T_z) = (\pm 1/2,\mp 1/2)$ (dark pink region). b) Critical current versus level position in the cotunneling limit. The critical current undergoes a $0$-$\pi$ transition when 
$\veps-\Delta_{SO}/2=0$. c) Discrete Josephson current  (in units of $\frac{e\Delta}{\hbar}$) versus $\phi$ in the Kondo limit. d) Discrete Josephson current for $\phi=\pi$ versus SO coupling. As the system changes from SU(4) to SU(2) Kondo symmetries, $I_J^{dis}$ goes from zero to maximum.}
\label{fig:su4su2}
\end{figure*}
The notation $E_{1(2)}$ indicates whether the Kramers doublet which contributes to the ABS is the ground (excited) state at $B_{||}=0$ (Fig. 1b). Importantly, each Kramers doublet gives two solutions in Eq. (\ref{eq:andreevbound}) so in general we obtain four ABS. The two outer (inner) solutions correspond to $E_{1(2)}$ (Fig. 1c). 
The results for the Josephson current are shown in Figs. 2a  and 2b where
a $0$-$\pi$ transition occurs for $B_\parallel\gtrsim B_c$. The transition can be understood by studying the ABS spectrum as a function of $\phi$ for different $B_\parallel$(Fig. 2c).  When $B_{||}\gtrsim B_c$, the two inner ABS cross at $E_F=0$.  Owing to this, the occupied ABS for $B_{||}\gtrsim B_c$ belong to the \emph{same} Kramers doublet (the one formed by ($K,\up$) and ($K',\down$) which are, of course, no longer degenerate). Importantly, they carry supercurrents of \emph{opposite} sign which leads to a negligible $I_{dis}$. The main contribution is thus given by the continuum part which results in $\pi$ behavior \cite{Spivak:91}. In Fig. 2d we plot the ABS as a function of gate voltage and different  $B_\parallel$. At zero magnetic field, the SO splitted ABS show a diamond-like shape, similarly to spin-slit ABS due to Coulomb Blockade \cite{Vecino:03,Pillet:10}. As $B_\parallel$ increases, the diamond closes and, ultimately, the two inner ABS become degenerate when $B_\parallel=B_c$. When  $B_{||}\gtrsim B_c$, the ABS cross at $E_F$. After the crossing, the occupied ABS belong to the same Kramers doublet for a large range of $|V_g|<\Delta$ resulting in a $\pi$ transition which is robust as the gate voltage is varied (Figs. 2e,2f).
%\paragraph{Finite $U$---}
We include the effect of the Coulomb repulsion by first considering the large gap limit, i.e., $\Delta\to\infty$, where the problem can be mapped onto an effective low-energy model ($U\ll \Delta$) with a superconducting pair potential due to the proximity effect $\Delta_D=\Gamma cos(\phi/2)$. Direct diagonalization produces results for the ground state energy $E_{GS}(\phi)$ and trivially $I_J=I_J^{dis}$ (in this limit, this is the only contribution to the Josephson current). 
Owing to SO, the total spin $S$ and orbital pseudospin $T$ are no good quantum numbers. Instead,
 $\calh_D$ has a  block diagonal form using the total projections $(S_z,T_z)$, with $\veps_{\tau,s}=\veps+1/2\tau s\Delta_{SO}$, as a basis. For $\phi=\pi$, we find the analytical solution
\begin{widetext}
\beq\label{groundstates}
E_{GS}(\phi=\pi) = 
\begin{cases}
\veps-\frac{1}{2} \Delta_{SO} & \text{for $U > -\veps + \frac{1}{2}\Delta_{SO}$, $(S_z,T_z) = (\pm 1/2,\mp 1/2)$} \\
2\veps - \Delta_{SO} + U & \text{for $-\frac{1}{2}\veps - \frac{1}{4} \Delta_{SO} < U < -\veps+ \frac{1}{2}\Delta_{SO}$, $(S_z,T_z) = (0,0)$,$(1,0)$,$(0,1)$ } \\
3\veps - \frac{1}{2}\Delta_{SO} + 3U & \text{for $-\frac{1}{3}\veps -\frac{1}{6}\Delta_{SO} < U < -\frac{1}{2}\veps - \frac{1}{4} \Delta_{SO}$, $(S_z,T_z) = (\pm 1/2,\pm 1/2)$} \\
4\veps + 6U & \text{for $U < -\frac{1}{3}\veps -\frac{1}{6}\Delta_{SO}$, $(S_z,T_z) = (0,0)$}
\end{cases}
\edq
\end{widetext}
The ground state for arbitrary $\phi$ has to be calculated numerically (Fig. 3a shows the phase diagram for $\phi=0$). Nevertheless, it can be shown (by comparing with the approximate boundaries obtained by perturbation theory in $\Delta_D$, red lines in Fig. 3a) that for large $U$ the ground state is always $(S_z,T_z) = (\pm 1/2,\mp 1/2)$ with energy $E_{GS}(\phi)=2\veps-1/2(\sqrt{4\Gamma^2cos^2(\phi/2)+(\Delta_{SO}+2\veps+3U)^2}+3U)$. While we cannot identify this state with a $\pi$ phase, it is likely that the inclusion of quantum fluctuations, by considering a finite gap, will stabilize the system towards this phase. Indeed, cotunneling corrections for ($\Gamma \ll \Delta$), present $\pi$ phases. This can be shown by employing second-order perturbation theory in $\Gamma$ (namely fourth-order cotunneling processes, see supplementary info) \cite{Novotn:05}. In this limit, we find a supercurrent $I_J = I_c \sin(\phi)$ such that the overall sign of $I_c$ governs the $0$ or $\pi$-character. In particular, the $0$-$\pi$-junction transition takes place at the value of $\veps$ corresponding 
to the resonant condition $\veps-\Delta_{SO}/2=E_F=0$, with a $\pi$ phase for $\veps<\Delta_{SO}/2$, such that the transition can be tuned by a gate voltage. Numerical results are shown in Fig. 3b.

Beyond cotunneling, higher order tunneling events lead to Kondo physics.  Here, we consider the large-$U$ limit (supplementary info) where
simultaneous fluctuations in the spin and orbital quantum numbers lead to a highly symmetric SU(4) Kondo effect (for a Kondo temperature $T_{K, SU(4)}>>\Delta)$. When $T_{K,SU(4)} \gg \Delta_{SO}$, we find
\begin{equation}
I_J^{dis}=\frac{e\Delta}{2\hbar}\sum_{\eta=\pm} \frac{\sin(\phi)}
{[(1+ \eta\alpha)^2+1][(1+ \eta\alpha)^2+cos^2({\phi\over 2})]}, 
\end{equation}
with $\alpha=\frac{\Delta_{SO}}{2T_{K,SU(4)}}$. When $T_{K,SU(4)} \ll \Delta_{SO}$,
only the lower dot  level participates in producing an SU(2) Kondo state. In the limit $T_{K,SU(2)} \gg \Delta$, the ABS are simply  $E_{1}= \pm\Delta\cos(\phi/2)$, namely the ABS of a single contact with unitary transmission. The corresponding supercurrent is $I_J^{dis} =\frac{e\Delta}{\hbar} \sin(\phi/2)$, with $|\phi|<\pi$ \cite{Beenakker:91}.  
Fig. 3c summarizes these results. For both symmetries, the Josephson current always exhibits a $0$-junction behavior but the magnitude strongly depends on $\Delta_{SO}$, as shown in Fig. 3d. For $\Delta_{SO}=0$, we recover the results of Ref. \cite{Zazunov:010}.

%\paragraph{Conclusion.---}
In closing, we have demonstrated that SO coupling induces a $0-\pi$ transition in the Josephson current through a QD NT coupled to superconducting leads. Our calculations, which cover all relevant transport regimes, non-interacting, Coulomb Blockade, cotunneling and Kondo, determine in a precise manner the conditions for the transition in terms of system parameters which can be tuned experimentally. Our predictions are relevant in view of recent experimental advances in transport through ultra-clean NTs with SO coupling \cite{Kuemmeth:08}. Furthermore, most of the physics discussed here is inherent to the rich behavior that ABS show in the presence of SO coupling. We therefore expect that tunneling spectroscopy of individual ABS, like in the experiments of Ref. \cite{Pillet:10}, may also reveal the effects described here. Microwave spectroscopy of excited ABS \cite{Bergeret:10} is one further experimental example where our findings may be tested.

\begin{acknowledgments} R.A. and R.L. acknowledge funding from MICINN Spain (Grants No. FIS2009-08744 and No. FIS2008-00781). 
\end{acknowledgments}
\appendix
\section{NT Model}
We consider a single wall NT whose low energies can be described by expanding the momentum near the Dirac points of graphene $\calh_0= \hbar v_F(k_y \tau_3 \otimes \sm_1 + k_x \tau_0 \otimes \sm_2)$, here $v_F$ is the Fermi velocity, $\tau_3$ is a Pauli matrix acting on isospin ($K,K'$) space (with eigenvalues $\tau=\pm 1$) whereas the Pauli matrices $\sm_1$ and $\sm_2$ act in sublattice space (the two carbon atoms in the primitive unit cell of the graphene honeycomb lattice). $k_x$ and $k_y$ are the momenta along the NT axis and circumferential direction, respectively. The eigenvalues of  $\calh_0$ are $E_0(k_x,k_y)=\pm \hbar v_F\sqrt{{k^2_x}+{k^2_y}}$. Imposing periodic boundary conditions, $k_y$ is quantized as $k_y=2\tau\nu/3D$ (lowest mode), where $D$ is the NT diameter and $\nu$ depends on the type of tube. In the following, we will consider small bandgap tubes parametrized as $k_y=\tau k_g$. We also include a magnetic field $B_{||}$ applied parallel to the NT axis. $B_{||}$ induces an Aharonov-Bohm flux $\Phi_{AB}=B_{||}\pi D^2$ such that $k_y=\tau k_g+\Phi_{AB}/D\Phi_0$, with $\Phi_0=h/e$ being the flux quantum. Besides this orbital shift, $B_{||}$  also induces the standard Zeeman shift in the spin sector $\calh_Z=\frac{1}{2} g\mu_B B_{\parallel} \tau_0 \otimes \sm_0 \otimes s_3$, with $s_3$ being a Pauli matrix (eigenvalues $s=\pm 1$) describing the spin projection along the tube axis. Finally, the SO coupling term has the form $\calh_{SO}=\left(\Delta^1_{SO} \tau_3 \otimes \sm_1 \otimes s_3 + \Delta^0_{SO} \tau_3 \otimes \sm_0 \otimes s_3 \right)$, which includes off-diagonal $\Delta^1_{SO}$  and diagonal $\Delta^0_{SO}$ \cite{Izumida:09,Jeong:09} terms in sublattice space . The eigenvalues of the full $\calh=\calh_0+\calh_Z+\calh_{SO}$ read $E_{s,\tau}(k_x,k_y)=\pm \hbar v_F\sqrt{{k^2_x}+{k^2_y}}+s(\tau\Delta^0_{SO}+\frac{1}{2} g\mu_B B_{\parallel})$, here $\Delta^1_{SO}$ has been absorbed in $k_y$ as $k_y=\tau k_g+\Phi_{AB}/D\Phi_0+s\Delta^1_{SO}/\hbar v_F$.  Finite intervalley scattering $\Delta_{K,K'}$ introduces anticrossings in the spectrum when spin polarized orbital states are degenerate (not shown). 
\subsection{QD Bound states} The total (low-energy) Hamiltonian for a quantum dot carbon nanotube with spin-orbit coupling can be written as \cite{Izumida:09,Jeong:09}
\begin{widetext}
\beq
\calh(x) = \hbar v_F(k_y \tau_3 \otimes \sm_1 + k_x \tau_0 \otimes \sm_2) \otimes s_0 + \left(\Delta_1 \tau_3 \otimes \sm_1 \otimes s_3 + \Delta_0 \tau_3 \otimes \sm_0 \otimes s_3 \right) + \frac{1}{2} g\mu_B B_{\parallel} \tau_0 \otimes \sm_0 \otimes s_3 + V(x)
\label{eq:soham}
\edq
\end{widetext}
Here $v_F$ is the Fermi velocity, $\tau_3$ is a Pauli matrix acting on isospin ($K,K'$) space (with eigenvalues $\tau=\pm 1$) whereas the Pauli matrices $\sm_1$ and $\sm_2$ act in sublattice space (they account for the two carbon atoms in the primitive unit cell of the honeycomb lattice describing graphene). $k_x$ and $k_y$ are the momenta along the NT axis and circumferential direction, respectively. The term $V(x)$ describes the potential induced by the electrostatic gates and is defined as a simple step potential of the form
\beq
V(x) =
\begin{cases}
V_0, & \text{$|x| > \ell$} \\
0,   & \text{$|x| < \ell$}
\end{cases}\nonumber
\edq
\begin{figure}
\centering
\includegraphics[width=0.4\textwidth]{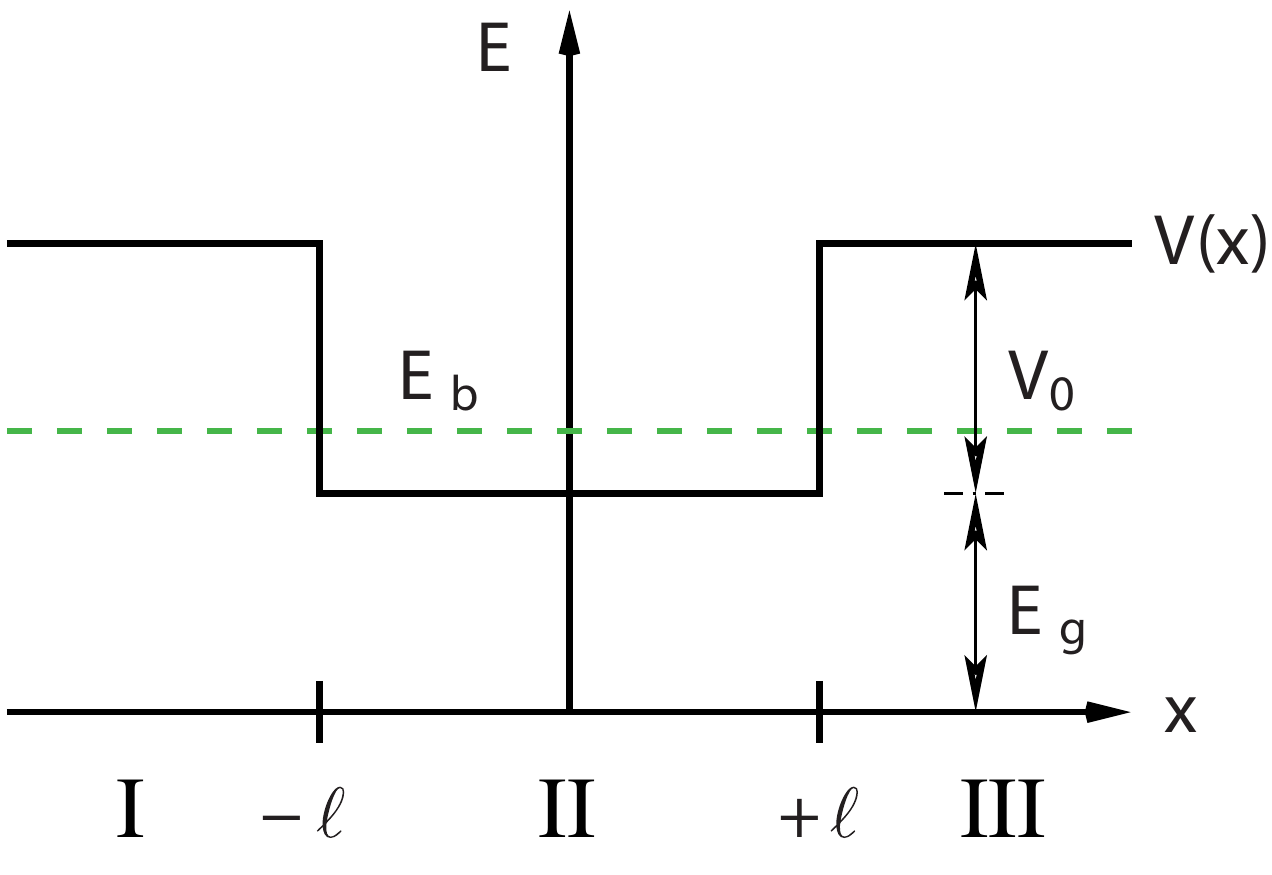}
\caption[Energy diagram of a CNT-QD]
{Energy diagram of a CNT-QD with electrostatic gates inducing a potential $V(x)$ along the nanotube axis.
Bound state energy $E$ (green dashed line) obeys a criterion $E_g < E_b < V_0 + E_g$.}
\label{fig:QDW}
\end{figure}
We use the following ansatz \cite{Bulaev:08,Weiss:010} for the electronic wavefunctions $\psi^{\tau}(t)$ defined in different intervals (see Fig. 1)
\bes
\begin{align}
\psi_{I}^{\tau}(x) &= A e^{q_x x} \binom{z_{k_y,-iq_x,c/v}^{\tau}}{1} \nonumber\\
\psi_{II}^{\tau}(x) &= B_1 e^{ik_x x} \binom{z_{k_y,k_x,c/v}^{\tau}}{1} + B_2 e^{-ik_x x} \binom{z_{k_y,-k_x,c/v}^{\tau}}{1} \nonumber\\
\psi_{III}^{\tau}(x) &= C e^{-q_x x} \binom{z_{k_y,iq_x,c/v}^{\tau}}{1}\nonumber
\end{align}
\label{eq:wf}
\eds
with momenta $k_x = \sqrt{\left(\frac{E}{\hbar v_F}\right)^2 - k_y^2}, \qquad q_x = \sqrt{k_y^2 - \left(\frac{E-V_0}{\hbar v_F}\right)^2}$ and  $k_y = \tau k_g + \frac{\Phi_{AB}}{R\Phi_0} + \frac{s\Delta_1}{\hbar v_F}~ (\text{for the lowest, $m=0$, subband}$). The corresponding energies are 
$\veps_{\tau,s}= \pm \frac{\hbar v_F}{L} \sqrt{(Lk_x)^2 + (Lk_y)^2} + \left( \Delta_0 \tau + \frac{1}{2} g\mu_B B_{\parallel} \right)s$, $E \equiv \veps_{\tau,s} - \left( \Delta_0 \tau + \frac{1}{2} g\mu_B B_{\parallel} \right)s$
and $z_{k_x,k_y,c/v}^{\tau} = \pm \frac{\tau k_y - ik_x}{\sqrt{k_x^2 + k_x^2}}$, with $L = 2\ell$. Here, the subscripts $c$ and $v$ correspond to conduction and valence bands, respectively. 
The energy levels $\veps_{\tau,s}$ are found from the continuity of the wavefunction $\psi^{\tau}(x)$ at all potential steps. That is,
\begin{widetext}
\bes
\begin{align}
A e^{-q_x \ell} \binom{z_{k_y,-iq_x,c/v}^{\tau}}{1} &= B_1 e^{-ik_x\ell}\binom{z_{k_y,k_x,c/v}^{\tau}}{1} + B_2 e^{ik_x\ell} \binom{z_{k_y,-k_x,c/v}^{\tau}}{1} \qquad \Leftarrow \qquad \psi_{I}^{\tau}(-\ell) = \psi_{II}^{\tau}(-\ell)
\label{eq:matchingI}
\\
C e^{-q_x\ell} \binom{z_{k_y,iq_x,c/v}^{\tau}}{1} &= B_1 e^{ik_x\ell}\binom{z_{k_y,k_x,c/v}^{\tau}}{1} + B_2 e^{-ik_x\ell}\binom{z_{k_y,-k_x,c/v}^{\tau}}{1} \qquad \Leftarrow \qquad \psi_{II}^{\tau}(\ell) = \psi_{III}^{\tau}(\ell)
\label{eq:matchingII}
\end{align}
\eds
Using Eqs.~\eqref{eq:matchingI}, we solve for $B_1$ and $B_2$ and get
\beq
\binom{B_1}{B_2} = \frac{\pm iAe^{-q_x\ell}}{2k_x/\sqrt{k_x^2+k_y^2}} 
\binom{e^{ik_x\ell} z_{k_y,-iq_x,c/v}^{\tau}-e^{ik_x\ell}z_{k_y,-k_x,c/v}^{\tau}}{-e^{-ik_x\ell} z_{k_y,-iq_x,c/v}^{\tau} + e^{-ik_x\ell}z_{k_y,k_x,c/v}^{\tau}}
\label{eq:bobt}
\edq
where $+(-)$ belongs to the conduction (valence) band.
Also, from Eq.~\eqref{eq:matchingII} we have
\beq
z_{k_y,iq_x,c/v}^{\tau} = \frac{B_1 e^{ik_x\ell}z_{k_y,k_x,c/v}^{\tau} + B_2 e^{-ik_x\ell}z_{k_y,-k_x,c/v}^{\tau}}{B_1 e^{ik_x\ell} + B_2 e^{-ik_x\ell}}
\label{eq:ebb}
\edq
Substituting Eq.~\eqref{eq:bobt} into Eq.~\eqref{eq:ebb} yields
\beq
\frac{\tau k_y+ q_x}{\sqrt{k_y^2 - q_x^2}} =
\frac{\frac{\tau k_y - q_x}{\sqrt{(k_y^2 - q_x^2)(k_x^2+k_y^2)}} (\tau k_y\sin 2k_x\ell - k_x\cos 2k_x\ell) - \sin 2k_x\ell}
{-\frac{1}{\sqrt{k_y^2 + k_x^2}} (\tau k_y\sin 2k_x\ell + k_x\cos 2k_x\ell) + \frac{\tau k_y -q_x}{\sqrt{k_y^2-q_x^2}}\sin 2k_x\ell}
\edq

Simplifying the above equation gives
\beq
\sin 2k_x\ell = \frac{1}{\sqrt{(k_y^2-q_x^2)(k_x^2+k_y^2)}} \left(k_y^2\sin 2k_x\ell + k_x q_x\cos 2k_x\ell \right)
\edq
which yields
\beq
\tan 2k_x\ell = \frac{k_x q_x}{\sqrt{(k_y^2-q_x^2)(k_x^2+k_y^2)} - k_y^2}
= \frac{k_x q_x}{|E-V_0||E|/(\hbar v_F)^2 - k_y^2}
\label{eq:bs}
\edq
Fig. 1b in the main text shows a typical energy spectrum for realistic NT parameters. In particular, we use SO coupling values $\Delta_0 = 0.26 \text{meV}$ and $\Delta_1 = 0.053 \text{meV}$ such that $\Delta_{SO} = 0.4156 \text{meV}$. Rest of parameters: $k_g = -0.09 \text{nm}^{-1}$, $L = 100$ nm, $V_0 = 3.95 \text{meV}$, $\hbar v_F = 526.57 \text{meV nm}$ .
\end{widetext}
\subsection{Including superconducting leads}
The QD CNT coupled to superconducting leads is modelled as an Anderson Hamiltonian coupled to s-wave superconductors with BCS density of states.
This Hamiltonian can be written in second quantization as $\calh = \calh_{C} + \calh_{D} + \calh_{T}$
where
\bes
\begin{align}
\calh_{C} &= \sum_{\alpha=L/R,k,\tau,s} \xi_{k} c_{\alpha k \tau s}^{\dag}c_{\alpha k \tau s}\nonumber\\ 
&- \sum_{\alpha,k,\tau} \left[ \Delta_{\alpha}e^{i\phi_{\alpha}} c_{\alpha k \tau \up}^{\dag}c_{\alpha \bk \btau \down}^{\dag} + h.c. \right] \\
\calh_{D} &= \sum_{\tau,s} \veps_{\tau s} d_{\tau s}^{\dag} d_{\tau s} + U \sum_{(\tau,s) \ne (\tau',s')} n_{\tau s}n_{\tau's'} \\
\calh_{T} &= \sum_{\alpha=L/R,k,\tau,s} \left( V_{\alpha} c_{\alpha k \tau s}^{\dag} d_{\tau s} + h.c.\right),
\end{align}
\label{eq:modelsc}
\eds 
where $c_{\alpha k \tau  s}^{\dag}$ 
creates an electron on lead $\alpha\in\{L,R\}$ with energy 
$\xi_{k}$ and with quantum numbers $k$, $s$, and $\tau$ corresponding to the wave-vector, 
spin and orbital degree of freedom, respectively.  $\Delta$ is the superconducting gap and $\phi=\phi_L-\phi_R$ is the phase difference. $d_{\tau s}$ is the operator that annihilates an electron on the dot 
with energy $\veps_{\tau s}$ (where the dependence on gate voltage is implicit). $U$ denotes the intra- and inter-orbital charging energy and 
$n_{\tau s}=d^\dag_{\tau s} d_{\tau s}$ represents the occupation operator for the dot levels. The last term describes tunneling by means of energy-independent tunneling 
amplitudes $V_\alpha$  leading to tunneling rates 
$\Gamma_{\alpha}=\pi|V_{\alpha}|^2\rho$  ($\rho$ is the contact density of states). 
\section{Calculation of Andreev bound states and the Josephson current by using the Green's functions technique}
A powerful technique to obtain the total Josephson current through the system described above is the Green's function method where all physical quantities can be written in terms of the Green's functions
\bes
\begin{align}
\calg_{d}^{r,a}(t,t') &= \dbraket{\hatd,\hatd^{\dag}}^{r,a}
= \mp i\Theta(\pm t \mp t') \nbraket{[\hatd(t),\hatd^{\dag}(t')]_+}\nonumber \\
\calg_{d}^{<}(t,t') &= \dbraket{\hatd,\hatd^{\dag}}^<
= i \nbraket{\hatd^{\dag}(t')\hatd(t)}
\nonumber\\
\calg_{d}^{>}(t,t') &= \dbraket{\hatd,\hatd^{\dag}}^>
= -i \nbraket{\hatd(t)\hatd^{\dag}(t')}
\end{align}
\eds
Owing to the superconducting pairing, these Green's functions are matrices containing anomalous components. In the following, we write these matrices using the following the Nambu bispinors
\beq
\hatc_{\alpha k} = 
\begin{bmatrix}
c_{\alpha k +\up} \\
c_{\alpha \bk -\down}^{\dag} \\
c_{\alpha k -\up} \\
c_{\alpha \bk +\down}^{\dag}
\end{bmatrix}
\quad
\text{and}
\quad
\hatd =
\begin{bmatrix}
d_{+\up} \\
d_{-\down}^{\dag} \\
d_{-\up} \\
d_{+\down}^{\dag},
\end{bmatrix}
\edq

\begin{widetext}
Using the standard relation $z \dbraket{A,B} + \dbraket{[\calh,A],B} = \nbraket{[A,B]_+}$, together with the following commutation relations
\bes
\begin{align}
[\calh,d_{\tau s}] &= -\veps_{\tau s} d_{\tau s} - U d_{\tau s} {\sum_{\tau',s'}}' n_{\tau' s'} - \sum_{\alpha,k} V_{\alpha} c_{\alpha k\tau s} \\
[\calh,d_{\btau \bs}^{\dag}] &= \veps_{\btau \bs} d_{\btau \bs}^{\dag} + U d_{\btau \bs}^{\dag} {\sum_{\btau',\bs'}}' n_{\btau' \bs'} + \sum_{\alpha,k} V_{\alpha} c_{\alpha \bk\btau \bs}^{\dag} \\
[\calh,c_{\alpha k\tau s}] &= -\xi_{k} c_{\alpha k\tau s} + \sgn(s) \Delta_{\alpha}^{i\phi_{\alpha}} c_{\alpha\bk\btau\bs}^{\dag} - V_{\alpha} d_{\tau s} \\
[\calh,c_{\alpha \bk\btau \bs}^{\dag}] &= \xi_{\bk} c_{\alpha \bk\btau \bs}^{\dag} - \sgn(\bs) \Delta_{\alpha}^{-i\phi_{\alpha}} c_{\alpha k\tau s} + V_{\alpha} d_{\btau \bs}^{\dag},
\end{align}
\eds
we obtain 
\bes
\begin{align}
\hatg_d^{r,-1} \dbraket{\hatd,\hatd^{\dag}}^r &= 1 + \sum_{\alpha,k}\hatV_{\alpha}\hsm_3 \dbraket{\hatc_{\alpha k},\hatd^{\dag}}^r + U \hsm_3 \dbraket{\begin{pmatrix} d_{+\up}\bn_{+\up} \\ \bn_{-\down} d_{-\down}^{\dag} \\ d_{-\up} \bn_{-\up} \\ \bn_{+\down} d_{+\down}^{\dag} \end{pmatrix},\hatd^{\dag}} \\
\hatg_{\alpha k}^{r,-1} \dbraket{\hatc_{\alpha k},\hatd^{\dag}}^r &= \hsm_3\hatV_{\alpha}\dbraket{\hatd,\hatd^{\dag}}^r
\end{align}
\label{eom}
\eds
where $\bn_{\tau s} = \sum_{\tau',s'}' n_{\tau' s'}$ (the prime in the summation means $(\tau',s') \ne (\tau,s)$) and 
\beq
\hatg_d^{r,-1} = 
\begin{pmatrix}
z - \veps_{+\up} &                    &                  &                    \\
                 & z + \veps_{-\down} &                  &                    \\
                 &                    & z - \veps_{-\up} &                    \\
                 &                    &                  & z + \veps_{+\down}
\end{pmatrix}
, \qquad
\hatg_{\alpha k}^{r,-1} = 
\begin{pmatrix}
z - \xi_{k}                         & \Delta_{\alpha} e^{i\phi_{\alpha}}   &                                      &                                    \\
\Delta_{\alpha} e^{-i\phi_{\alpha}} & z + \xi_{\bk}                        &                                      &                                    \\
                                    &                                      & z - \xi_{k}                          & \Delta_{\alpha} e^{i\phi_{\alpha}} \\
                                    &                                      & \Delta_{\alpha} e^{-i\phi_{\alpha}} & z + \xi_{\bk}
\end{pmatrix}
\edq
and
\beq
\hatV_{\alpha} =
\begin{pmatrix}
V_{\alpha} &            &            &           \\
           & V_{\alpha} &            &           \\
           &            & V_{\alpha} &           \\
           &            &            & V_{\alpha}
\end{pmatrix}
, \qquad
\hsm_3 = 
\begin{pmatrix}
1 &    &   &    \\
  & -1 &   &    \\
  &    & 1 &    \\
  &    &   & -1
\end{pmatrix}
\edq
\end{widetext}
Owing to the presence of Coulomb Interactions, $U\neq 0$, the equations of motion for the Green's functions, Eqs. (\ref{eom}), cannot be closed and we need some approximations which we discuss next.
\subsection{Non-interacting limit $U=0$}
\subsubsection{Retarded Green's functions}
In the noninteracting case, $U = 0$, the equations of motions can be closed such that analytical expressions for the Green's functions and self-energies can be obtained. In particular, the retarded ones read: 
\beq
\dbraket{\hatd,\hatd^{\dag}}^r = \left[ \hatg_d^{r,-1} - \hatsig_0^r \right]^{-1}
\label{eq:nonham}
\edq
where
\begin{widetext}
\beq
\begin{split}
\hatsig_0^r &= \sum_{\alpha,k} \hatV \hsm_3 \hatg_{\alpha k}^{r} \hsm_3 \hatV 
%&
= -\Gamma 
\begin{pmatrix} 
\beta_d & \beta_o \cos(\phi/2) & & \\ 
\beta_o \cos(\phi/2) & \beta_d & & \\
 & & \beta_d & \beta_o \cos(\phi/2) \\ 
 & & \beta_o \cos(\phi/2) & \beta_d
\end{pmatrix}
\end{split}
\edq
\end{widetext}
with $\Gamma = 2\pi\rho_N(0) V^2$ and
\bes
\begin{align}
\beta_d &= 
\begin{cases}
\frac{\omega}{\sqrt{\Delta_{\alpha}^2 - \omega^2}} & \text{if $|\omega| < \Delta$}    \\
i\frac{|\omega|}{\sqrt{\omega^2 - \Delta_{\alpha}^2}} & \text{if $|\omega| > \Delta$} 
\end{cases}
\\
\beta_o &= 
\begin{cases}
\frac{\Delta_{\alpha}}{\sqrt{\Delta_{\alpha}^2 - \omega^2}} & \text{if $|\omega| < \Delta$}    \\
i\frac{\sgn(\omega)\Delta_{\alpha}}{\sqrt{\omega^2 - \Delta_{\alpha}^2}} & \text{if $|\omega| > \Delta$} 
\end{cases}
\end{align}
\eds
Using $\Delta_L = \Delta_R \equiv \Delta$, $V_L = V_R = V$, and $\phi_L = -\phi_R = \phi/2$, the retarded Green's function reads:
\begin{widetext}
\beq
\dbraket{\hatd,\hatd^{\dag}}^r = 
\begin{pmatrix}
\frac{1}{D_+} 
\begin{pmatrix}
z+\veps_{-\down}+\Gamma\beta_d & -\Gamma\beta_o\cos(\phi/2) \\
-\Gamma\beta_o\cos(\phi/2) & z-\veps_{+\up}+\Gamma\beta_d
\end{pmatrix}
& \bf{0}
\\
\bf{0}
& \frac{1}{D_-} 
\begin{pmatrix}
z+\veps_{+\down}+\Gamma\beta_d & -\Gamma\beta_o\cos(\phi/2) \\
-\Gamma\beta_o\cos(\phi/2) & z-\veps_{-\up}+\Gamma\beta_d
\end{pmatrix}
\end{pmatrix}
\edq
where
\bes
\begin{align}
D_+ &= (z-\veps_{+\up}+\Gamma\beta_d)(z+\veps_{-\down}+\Gamma\beta_d)-(\Gamma\beta_o\cos(\phi/2))^2
\\
D_- &= (z-\veps_{-\up}+\Gamma\beta_d)(z+\veps_{+\down}+\Gamma\beta_d)-(\Gamma\beta_o\cos(\phi/2))^2.
\end{align}
\label{eq:dpdm}
\eds
\end{widetext}

\begin{widetext}
\subsubsection{Andreev Bound States}
When $|\omega| < \Delta$, the Andreev bound states can be determined from the poles of the Green's function. 
Namely, we just need to solve the determinant equation $Det[\calg_{d}^{r}(\omega)^{-1}]=0$. Using Eq.~\eqref{eq:nonham} we obtain the following equation:
\beq
\left|\hatg_d^{r,-1}(\omega) - \hatsig_0^r(\omega) \right| = D_+ D_- = 0 
\edq
Explicitly,
\beq
\left[(\omega - \veps_{+\up} + \Gamma\beta_d)(\omega+\veps_{-\down}+\Gamma\beta_d) - \left(\Gamma\beta_o\cos(\phi/2)\right)^2\right]
\left[(\omega - \veps_{-\up} + \Gamma\beta_d)(\omega+\veps_{+\down}+\Gamma\beta_d) - \left(\Gamma\beta_o\cos(\phi/2)\right)^2\right]
= 0.
\edq
At this point it is important to note the full equivalence of the Green's function method with the Bogoliubov-DeGennes Hamiltonian method (indeed, the Green's function has precisely Bogoliubov-DeGennes structure).
The Andreev bound states give rise to delta-function contributions in the spectral density. The weights can be found by the residues of the Green's function at these poles. Explicitly,
\bes
\begin{align}
\Im\left[\dbraket{\hatd,\hatd^{\dag}}\right]_{11} &= -\pi\sum_{E_{2}} Z_{b+} \delta(\omega-E_{2}) \\
\Im\left[\dbraket{\hatd,\hatd^{\dag}}\right]_{12} &= -\pi\sum_{E_{2}} Q_{b+} \delta(\omega-E_{2}) \\
\Im\left[\dbraket{\hatd,\hatd^{\dag}}\right]_{33} &= -\pi\sum_{E_{1}} Z_{b-} \delta(\omega-E_{1}) \\
\Im\left[\dbraket{\hatd,\hatd^{\dag}}\right]_{34} &= -\pi\sum_{E_{1}} Q_{b-} \delta(\omega-E_{1})
\end{align}
\eds
where
\bes
\begin{align}
Z_{b\pm} &= \lim_{\omega\to E_{2(1)}} \frac{\omega + \veps_{\mp\down} - \hatsig_{0,22(44)}^r(\omega)}{D_{\pm}'(\omega)} 
\label{eq:zweights}
\\
Q_{b\pm} &= \lim_{\omega\to E_{2(1)}} \frac{\hatsig_{0,12(34)}^r(\omega)}{D_{\pm}'(\omega)}
\label{eq:qweights}
\end{align}
\eds
and
\bes
\beq
\left(E_{2(1)} - \veps_{\pm\up} + \frac{\Gamma E_{2(1)}}{\sqrt{\Delta^2-E_{2(1)}^2}}\right)
\left(E_{2(1)} + \veps_{\mp\down} + \frac{\Gamma E_{2(1)}}{\sqrt{\Delta^2-E_{2(1)}^2}}\right)
- \frac{\Gamma^2\Delta^2\cos^2(\phi/2)}{\Delta^2-E_{2(1)}^2} 
= 0
\label{eq:andreevbound-appendix}
\edq
\beq
D_{\pm}'(\omega) = \left(1 + \frac{\Gamma\Delta^2}{(\Delta^2-\omega^2)\sqrt{\Delta^2-\omega^2}}\right)
\left(2\omega - \veps_{\pm\up} + \veps_{\mp\down} + \frac{2\Gamma\omega}{\sqrt{\Delta^2 - \omega^2}}\right)
- \frac{2\omega\Gamma^2\Delta^2\cos^2(\phi/2)}{(\Delta^2 - \omega^2)^2}.
\edq
\eds
Eq. (\ref{eq:andreevbound-appendix}) corresponds to Eq. (1) in the main text.
\end{widetext}
\subsubsection{Josephson Current}
The current through a given lead $\alpha$ can be written as $I_{\alpha}^{dc} = e\frac{d\nbraket{n_{\alpha}}}{dt}$ with $n_{\alpha} = \sum_{k\in \alpha,\tau,s} c_{\alpha k\tau s}^{\dag} c_{\alpha k\tau s}$. Owing to the Josephson effect, this expression contains a dissipationless component (nonzero current at zero bias voltage) when there is a superconducting phase difference. Thus, the Josephson current can be extracted from the general $I_{\alpha}^{dc}$ by just studying the limit of zero applied bias voltage
$I_{\alpha}\equiv I_{\alpha}^{dc}|_{V_{dc}=0}= e\frac{d\nbraket{n_{\alpha}}}{dt}|_{V_{dc}=0}
= \frac{2e}{\hbar} 
\Re\left[ \sum_{k\in \alpha} \tr \left(\hatV_{\alpha} \dbraket{\hatc_{\alpha k},\hatd^{\dag}}^<(t,t) \right) \right]|_{V_{dc}=0}$. Using the nonequilibrium Green's function and the equation of motion methods, one finds that the Josephson current can be expressed as
\beq
I_{\alpha} = \frac{2e}{\hbar} \Re \int \frac{d\omega}{2\pi}~ \tr \left[\hsm_3\left(\brsig_0^< \dbraket{\hatd,\hatd^{\dag}}^a + \brsig_0^r \dbraket{\hatd,\hatd^{\dag}}^<\right)\right]
\edq
where
\begin{widetext}
\bes
\beq
\begin{split}
\brsig_0^r &= \sum_{\alpha,k} \hatV \hsm_3 \hatg_{\alpha k}^{r} \hsm_3 \hatV 
%&
= -\frac{\Gamma}{2} 
\begin{pmatrix} 
\beta_d & \beta_o e^{+i\phi_{\alpha}} & & \\ 
\beta_o e^{-i\phi_{\alpha}} & \beta_d & & \\
 & & \beta_d & \beta_o e^{+i\phi_{\alpha}} \\ 
 & & \beta_o e^{-i\phi_{\alpha}} & \beta_d
\end{pmatrix}
\end{split}
\edq
\beq
\begin{split}
\brsig_0^< &= \sum_{k} \hatV \hsm_3 \hatg_{\alpha k}^{<} \hsm_3 \hatV 
%&
= \Gamma \Theta(|\omega|-\Delta) f(\omega)
\begin{pmatrix} 
\beta_d & \beta_o e^{+i\phi_{\alpha}} & & \\ 
\beta_o e^{-i\phi_{\alpha}} & \beta_d & & \\
 & & \beta_d & \beta_o e^{+i\phi_{\alpha}} \\ 
 & & \beta_o e^{-i\phi_{\alpha}} & \beta_d
\end{pmatrix}
\end{split}
\edq
\eds
with $\Gamma = 2\pi\rho_N(0) V^2$ and  
\bes
\begin{align}
\beta_d &= 
\begin{cases}
\frac{\omega}{\sqrt{\Delta_{\alpha}^2 - \omega^2}} & \text{if $|\omega| < \Delta$}    \\
i\frac{|\omega|}{\sqrt{\omega^2 - \Delta_{\alpha}^2}} & \text{if $|\omega| > \Delta$} 
\end{cases}
\\
\beta_o &= 
\begin{cases}
\frac{\Delta_{\alpha}}{\sqrt{\Delta_{\alpha}^2 - \omega^2}} & \text{if $|\omega| < \Delta$}    \\
i\frac{\sgn(\omega)\Delta_{\alpha}}{\sqrt{\omega^2 - \Delta_{\alpha}^2}} & \text{if $|\omega| > \Delta$}.
\end{cases}
\end{align}
\eds
One important advantage of this method is that the Josephson current can be easily split into two parts $I = I_{dis} + I_{con}$. The first part is the so-called discrete contribution and corresponds to the Josephson current carried by Andreev Bound states. The second term, the so-called continuous part $I_{con}$, corresponds to the current given by the continuous spectrum of states above the gap. 
Both expressions can be written analytically as:
\bes
\begin{align}
I_{dis} &= -\frac{2e\Gamma}{\hbar} 
\int \frac{d\omega}{2\pi} \Theta(\Delta - |\omega|) f(\omega) \frac{\Delta}{\sqrt{\Delta^2 - \omega^2}} 
\sin(\phi/2) \Im\left[\dbraket{\hatd,\hatd^{\dag}}_{21}^r+\dbraket{\hatd,\hatd^{\dag}}_{12}^r + \dbraket{\hatd,\hatd^{\dag}}_{43}^r+\dbraket{\hatd,\hatd^{\dag}}_{34}^r\right]
\\
I_{con} &= -\frac{2e\Gamma}{\hbar} 
\int \frac{d\omega}{2\pi}~ \Theta(|\omega|-\Delta)f(\omega) \frac{\sgn(\omega)\Delta}{\sqrt{\omega^2 - \Delta^2}}
\sin(\phi/2) \Re\left[\dbraket{\hatd,\hatd^{\dag}}_{21}^r+\dbraket{\hatd,\hatd^{\dag}}_{12}^r + \dbraket{\hatd,\hatd^{\dag}}_{43}^r+\dbraket{\hatd,\hatd^{\dag}}_{34}^r\right],
\end{align}
\eds
where, again, $\Delta_L = \Delta_R \equiv \Delta$, $V_L = V_R = V$, and $\phi_L = -\phi_R = \phi/2$.
Explicitly,
\bes
\begin{align}
I_{dis} &= -\frac{e\Gamma^2}{\hbar} \sin(\phi) \left[ \sum_{E_{2}} \frac{f(E_{2})\Delta^2}{(\Delta^2-E_{2}^2)D_+'(E_{2})} + \sum_{E_{1}} \frac{f(E_{1})\Delta^2}{(\Delta^2-E_{1}^2)D_-'(E_{1})} \right]
\label{eq:Idisformula}
\\
I_{con} &= -\frac{e\Gamma^2}{\pi\hbar} \sin(\phi) \int d\omega~ \Theta(|\omega|-\Delta) \frac{f(\omega)\Delta^2}{(\omega^2 - \Delta^2)} \Im \left[ \frac{1}{D_+(\omega)} + \frac{1}{D_-(\omega)}\right]
\end{align}
\eds

After some algebra, the discrete contribution can be rewritten as
\beq
I_{dis} = \frac{2e}{\hbar} \sum_{E_{1(2)}} f(E_{1(2)}) \frac{\partial E_{1(2)}}{\partial \phi},
\label{eq:Idisab}
\edq
which is the expression discussed in the main text.
\subsection{Cotunneling regime}
Expressions in the cotunneling regime can be obtained by lowest (second order) perturbation theory in $\Gamma$ \cite{Novotn:05}. Starting from the expression for the current
\beq
I = I_{\alpha} = -\frac{ie}{\hbar} \sum_{k\in \alpha,\tau,s} V_{\alpha} \nbraket{\left[ c_{\alpha k\tau s}^{\dag}d_{\tau s} - d_{\tau s}^{\dag} c_{\alpha k\tau s} \right]}
= \frac{2e}{\hbar} \sum_{k,\tau,s} \Im \left[ \nbraket{\calh_{T\alpha}^{-}} \right]
\edq
where
\beq
\calh_{T\alpha}^{-} = \sum_{\tau,s} \calh_{T\alpha\tau s}^{-}, \qquad  \calh_{T\alpha\tau s}^{-} = \sum_{k} V_{\alpha} c_{\alpha k\tau s}^{\dag} d_{\tau s}, \qquad \calh_{T\alpha}^{+} = (\calh_{T\alpha}^{-})^{\ast},
\edq
we perform a standard thermodynamic perturbation expansion in the tunneling and obtain the Josephson current in the lowest non-vanishing order (fourth order in $\calh_T$) as
\beq
I_{\alpha} = -\frac{2e}{\hbar} \Im \left[ \frac{1}{3!} \int_0^{\beta} d\tau_1 \int_0^{\beta} d\tau_2 \int_0^{\beta} d\tau_3 \nbraket{T_{\tau}\left(\calh_T(\tau_1)\calh_T(\tau_2)\calh_T(\tau_3)\calh_{T\alpha}^{-}\right)}_0 \right]
\edq

The Josephson current must involve two $\calh_T^+$ and two $\calh_T^-$, which can be chosen in three ways, and hence
\beq
I_{\alpha} = -\frac{e}{\hbar} \Im \left[ \int_0^{\beta} d\tau_1 \int_0^{\beta} d\tau_2 \int_0^{\beta} d\tau_3 \nbraket{T_{\tau}\left(\calh_{T\balpha}^+(\tau_1)\calh_{T\balpha}^+(\tau_2)\calh_{T\alpha}^-(\tau_3)\calh_{T\alpha}^{-}\right)}_0 \right]
\edq
where we have used that in order to have Cooper pair tunneling, the $\calh_T^+$ must belong to the opposite junction.
Next, if we choose the valley and spin of the last $\calh_T^-$ as, say, $(+,\up)$, it then means that the other $\calh_T^-$ carries $(-,\down)$. 
In the same way, the valley and spin of the two $\calh_T^+$ can be chosen. All in all, we thus obtain
\beq
I_{\alpha} = -\frac{e}{\hbar} \sum_{\tau,s} \sum_{\tau',s'} 
\Im \left[ \int_0^{\beta} d\tau_1 \int_0^{\beta} d\tau_2 \int_0^{\beta} d\tau_3 \nbraket{T_{\tau}\left(\calh_{T\balpha\btau'\bs'}^+(\tau_1)\calh_{T\balpha\tau' s'}^+(\tau_2)\calh_{T\alpha\btau\bs}^-(\tau_3)\calh_{T\alpha\tau s}^{-}\right)}_0 \right]
\label{eq:IJgds}
\edq
At arbitrary $B_{\parallel}$, the Josephson current can be written as $I_\alpha= I_c \sin(\phi)$, where the critical current reads
\beq
I_c = \frac{e\Gamma^2}{2\hbar\pi^2} \left[N(\veps_{+\up}) + N(\veps_{+\down}) + N(\veps_{-\up}) + N(\veps_{-\down})\right]
\edq
where
\beq
N(\veps) = -\int_{\Delta}^{\infty} dE \int_{\Delta}^{\infty} dE'~ \frac{\Delta}{\sqrt{E^2-\Delta^2}} \frac{\Delta}{\sqrt{E'^2-\Delta^2}} C(E,E',\veps,\bar{\veps})
\label{puro}
\edq
and
\begin{multline}
C(E,E',\veps,\bar{\veps}) =
\frac{e^{-\beta\veps}}{1 + \sum_{\tau,s} e^{-\beta\veps_{\tau s}}} 
\int_0^{\beta} d\tau_1 \int_0^{\beta} d\tau_2 \int_0^{\beta} d\tau_3~ \calf_L(E,\tau_3) \calf_R(E',\tau_1-\tau_2) \Theta(\tau_2-\tau_3)\Theta(\tau_3-\tau_1) e^{\veps \tau_2} e^{\bar{\veps}(\tau_3-\tau_1)}
\end{multline}
with $\bar{\veps}_{\tau s} = \veps_{\btau\bs}$. 
\end{widetext}
The functions $\calf_\alpha$ are related to the the anomalous Green's functions of the leads, which are defined as
\bes
\begin{align}
\scrf_{\alpha k+}(\tau,\tau') &= -\nbraket{T_{\tau} c_{\alpha \bk-\down}^{\dag}(\tau) c_{\alpha k+\up}^{\dag}(\tau')} \\
\scrf_{\alpha k-}(\tau,\tau') &= -\nbraket{T_{\tau} c_{\alpha \bk+\down}^{\dag}(\tau) c_{\alpha k-\up}^{\dag}(\tau')}
\end{align}
\eds
and given by
\bes
\begin{align}
\scrf_{\alpha k+}(\tau,\tau') &= \scrf_{\alpha k-}(\tau,\tau') \equiv \scrf_{\alpha k}(\tau,\tau') \nonumber\\
&= \frac{\Delta_{\alpha}e^{-i\phi_{\alpha}}}{2E_{\alpha k}} \calf_{\alpha}(E_{\alpha k},\tau-\tau')
\end{align}
\eds
where
\beq
\calf_{\alpha}(E_{\alpha k},\tau) \equiv e^{-E_{\alpha k}|\tau|} - 2\cosh\left(E_{\alpha k}\tau\right) n_F(E_{\alpha k})
\edq
Throughout, we assume low temperatures such that $\Delta_{L/R}\beta \gg 1$, and we thus approximate
\beq
\calf_{\alpha}(E_{\alpha k},\tau) \approx e^{-E_{\alpha k}|\tau|} - e^{-E_{\alpha k}(\beta-|\tau|)}
\edq
Here, first performing the imaginary time integration and then taking the approximation $\exp[-\beta E] \approx 0$, the function $C(E,E',\veps,\bar{\veps})$ is given by
\begin{widetext}
\begin{multline}
C(E,E',\veps,\bar{\veps}) = \frac{e^{-\beta\veps}}{1 + \sum_{\tau,s}e^{-\beta\veps_{\tau s}}} \left[ -\frac{e^{\beta\veps}}{(E+E')(E+\veps)(E'+\bar{\veps})}-\frac{e^{\beta\veps}}{(E+E')(E+\bar{\veps})(E'+\veps)} \right. \\
\left. + \frac{1}{(E+E'-\veps+\bar{\veps})(E-\veps)(E'-\veps)} + \frac{e^{\beta(\veps-\bar{\veps})}}{(E+E'+\veps-\bar{\veps})(E-\bar{\veps})(E'-\bar{\veps})}\right].
\end{multline}
\end{widetext}
In general, the integrals in Eq. (\ref{puro}) have to be evaluated numerically. For example, let us assume that all levels are well below the Fermi level. Then, at $T= 0$ only the lowest level contributes so that the critical current is given by
\begin{widetext}
\bes
\begin{align}
I_c = \frac{e\Gamma^2}{2\hbar\pi^2} \left[N(\veps_{+\down}) + N(\veps_{-\up})\right]
= 
\begin{cases} 
-\frac{e\Gamma^2}{\hbar\pi^2\Delta} \mathcal{M}_2(-\veps_{+\down}/\Delta) & \text{for $B_{\parallel} > 0$} \\ 
-\frac{e\Gamma^2}{\hbar\pi^2\Delta} \mathcal{M}_2(-\veps_{-\up}/\Delta) & \text{for $B_{\parallel} < 0$} 
\end{cases}
\end{align}
\eds
where
\beq
\mathcal{M}_2(\veps) = \int_{1}^{\infty} du \int_{1}^{\infty} dv~ \frac{1}{\sqrt{u^2-1}} \frac{1}{\sqrt{v^2-1}} \frac{1}{(u+v+\veps+\bar{\veps})(u+\veps)(v+\veps)}
\edq
Even in this simple case, the integral cannot be solved analytically. This is in contrast with the limit $B_{\parallel}=0$, where further analytical progress can be made. Assuming, for simplicity, $\veps_{+\up} = \veps_{-\down} = \veps_d + \Delta_{SO}/2$ and $\veps_{-\up} = \veps_{+\down} = \veps_d - \Delta_{SO}/2$,
the $0$ or $\pi$ character of the junction can be extracted by the overall sign of the critical current which reads $I_c=\frac{e\Gamma^2}{\hbar \pi^2} \left[N(\veps_{+\up}) + N(\veps_{-\up})\right]$
with 
\bes
\beq
N(\veps_{+\up}) = 
\begin{cases}
2 \mathcal{M}(\veps_{+\up}/\Delta)/\Delta & \text{for $\veps_d > +\Delta_{SO}/2$} \\
0 & \text{for $-\Delta_{SO}/2 < \veps_d < +\Delta_{SO}/2$} \\
0 & \text{for $\veps_{d} < -\Delta_{SO}/2$}
\end{cases}
\edq
and
\beq
N(\veps_{-\up}) = 
\begin{cases}
2\mathcal{M}(\veps_{-\up}/\Delta)/\Delta & \text{for $\veps_d > +\Delta_{SO}/2$} \\
-\mathcal{M}(-\veps_{-\up}/\Delta)/\Delta & \text{for $-\Delta_{SO}/2 < \veps_d < +\Delta_{SO}/2$} \\
-\mathcal{M}(-\veps_{-\up}/\Delta)/\Delta & \text{for $\veps_{d} < -\Delta_{SO}/2$}
\end{cases}
\edq
\eds
Here, the dimensionless function $\mathcal{M}(x)$ defined as $(x > - 1)$
\beq
\mathcal{M}(x) = \int_{1}^{\infty} du \int_{1}^{\infty} dv~ \frac{1}{\sqrt{u^2-1}}\frac{1}{\sqrt{v^2-1}}\frac{1}{(u+v)(u+x)(v+x)}
\edq
\end{widetext}
can be expressed by
\beq
\mathcal{M}(x) = \frac{(\pi/2)^2(1-x) - \arccos^2 x}{x(1-x^2)}, \qquad \text{with $x > -1$}
\edq
where the analytic continuations of $\arccos x = i\ln(x + \sqrt{x^2-1})$ for $x > 1$ is understood.
The function $\mathcal{M}(x)$ is always positive, it diverges at $x \to -1^+$, and then smoothly decays for increasing $x$ with the asymptote $\mathcal{M}(x) \sim \pi^2/4x^2$ for $x \to \infty$, which allows to extract analytical boundaries for the $0-\pi$ transition. In particular, we can establish the following criteria
\begin{enumerate}
\item If both levels $\veps_{+\up}$ and $\veps_{-\up}$ are above the Fermi level, the Josephson current is positive, i.e., 0-junction.
\item If both levels are below the Fermi level, it shows a $\pi$-junction behavior.
\item If the level $\veps_{+\up}$ is above the Fermi level and the other level $\veps_{-\up}$ is below the Fermi level, it is again a $\pi$-junction.
\end{enumerate}

For $\Delta_{SO} = 0$, the function $N(\veps_d)$ can be written as
\beq
N(\veps_{d}) = 
\begin{cases}
2\mathcal{M}(\veps_{-\up}/\Delta)/\Delta & \text{for $\veps_d > 0$} \\
-\frac{1}{2}\mathcal{M}(-\veps_{-\up}/\Delta)/\Delta & \text{for $\veps_{d} < 0$}
\end{cases}
\edq
an the critical current is given by $I_c = \frac{2e\Gamma^2}{\hbar\pi^2\Delta}N(\veps_d) = \frac{e\Gamma^2}{\hbar\pi^2\Delta} \left[4\Theta(\veps_d) - \Theta(-\veps_d)\right]\mathcal{M}(|\veps_d|/\Delta)$,
in agreement with Ref. \cite{Zazunov:010}.
\subsection{Kondo regime}
We study the Kondo regime in the large-$U$ limit by means of the slave boson method. Using this language, Eq.~\eqref{eq:modelsc} can be rewritten as
\begin{widetext}
\begin{multline}
\calh_{SB} = \sum_{\alpha=L/R,k,\tau,s} \xi_{k} c_{\alpha k \tau s}^{\dag}c_{\alpha k \tau s} 
- \sum_{\alpha,k,\tau} \left( \Delta_{\alpha}e^{i\phi_{\alpha}} c_{\alpha k\tau\up}^{\dag}c_{\alpha \bk \btau\down}^{\dag} + h.c. \right) \\
+ \sum_{\tau,s} \veps_{\tau s} f_{\tau s}^{\dag} f_{\tau s} 
+ \frac{1}{\sqrt{N}}\sum_{\alpha=L/R,k,\tau,s} \left( V_{\alpha} c_{\alpha k \tau s}^{\dag} b^{\dag} f_{\tau s} + h.c.\right)
+ \Lambda \left(\sum_{\tau,s} f_{\tau s}^{\dag}f_{\tau s} + b^{\dag}b - 1\right),
\label{eq:hsb}
\end{multline}
\end{widetext}
where  the physical fermionic operator is written as $d_{\tau s}^{\dag} = f_{\tau s}^{\dag} b$. The pseudofermion operator $ f_{\tau s}^{\dag}$ creates a state with spin $s$ and isospin $\tau$ and the slave boson operator $b$ annihilates an empty state. It can be shown that this mapping is exact provided that the constraint
\beq
\sum_{\tau,s} f_{\tau s}^{\dag}f_{\tau s} + b^{\dag}b = 1
\label{eq:constraintsone}
\edq
is fulfilled ($\Lambda$ is a Lagrange multiplier which enforces this constraint).
Note that the hybridization element is rescaled into $V_{\alpha} \to V_{\alpha}/\sqrt{N}$.
From the equation of motion for the boson field $b^{\dag}$, we have
\beq
\partial_t b^{\dag} = \frac{i}{\hbar} [\calh_{SB},b^{\dag}]
= \frac{i}{\hbar} \left[ \frac{1}{\sqrt{N}}\sum_{\alpha,k,\tau,s} V_{\alpha} f_{\tau s}^{\dag} c_{\alpha k\tau s} + \Lambda b^{\dag} \right]
\edq
In order to obtain self-consistent equations, we replace $\nbraket{b^{\dag}}$ by $\sqrt{N}\wbb^{\ast}$ and obtain
\beq
\frac{1}{N}\sum_{\alpha,k} V_{\alpha}\wbb \nbraket{\hatf^{\dag}(t) \hsm_3 \hatc_{\alpha k}(t)} + \Lambda |\wbb|^2 = 0
\label{eq:seone}
\edq
Eq.~\eqref{eq:seone} constitutes a set of self-consistent equations together with the constraint
\beq
\frac{1}{N} \nbraket{\hatf^{\dag}(t)\hsm_3 \hatf(t)} +  |\wbb|^2 = \frac{1}{N}
\edq
This mean field approximation, which neglects charge fluctuations, is reliable in the deep Kondo regime where only spin fluctuations are relevant.
\begin{widetext}
In the frequency space, the equations become
\bes
\begin{align}
\frac{1}{N} \sum_{\alpha,k} 
\int \frac{d\omega}{2\pi i} \tr \left[\wV_{\alpha}\hsm_3 \dbraket{\hatc_{\alpha k},\hatf^{\dag}}_{\omega}^<\right] + \Lambda \frac{\wGamma}{\Gamma} &= 0 \\
\frac{1}{N} \int \frac{d\omega}{2\pi i} \tr \left[\dbraket{\hatf,\hatf^{\dag}}_{\omega}^< \hsm_3 \right] + \frac{\wGamma}{\Gamma} &= \frac{1}{N}
\end{align}
\label{eq:sbse}
\eds

where
\beq
\wV_{\alpha} = \begin{pmatrix} V_{\alpha}\wbb & & & \\ & V_{\alpha}\wbb & & \\ & & V_{\alpha}\wbb & \\ & & & V_{\alpha}\wbb \end{pmatrix}
\edq
and $\wGamma = \Gamma\left|\wbb\right|^2$. At this point, we have to calculate the lesser Green's functions. To do that, we note that the mean-field Hamiltonian is given by
\begin{multline}
\calh_{MF} = \sum_{\alpha=L/R,k,\tau,s} \xi_{k} c_{\alpha k \tau s}^{\dag}c_{\alpha k \tau s} 
- \sum_{\alpha,k,\tau} \left( \Delta_{\alpha}e^{i\phi_{\alpha}} c_{\alpha k\tau\up}^{\dag}c_{\alpha \bk \btau\down}^{\dag} + h.c. \right) \\
+ \sum_{\tau,s} \weps_{\tau s} f_{\tau s}^{\dag} f_{\tau s} 
+ \sum_{\alpha=L/R,k,\tau,s} \left( V_{\alpha}\wbb^{\ast} c_{\alpha k \tau s}^{\dag} f_{\tau s} + h.c.\right)
+ \Lambda \left(|\wbb|^2 - 1\right)
\label{eq:mfham}
\end{multline}
where $\weps_{\tau s} = \veps_{\tau s} + \Lambda$.
The retarded Green's functions can be then written as
\beq
\dbraket{\hatf,\hatf^{\dag}}^r = \left[ \wg_f^{r,-1} - \wsig_0^r \right]^{-1}
\edq
where
\beq
\wg_f^{r,-1} = 
\begin{pmatrix}
z - \weps_{+\up} &                    &                  &                    \\
                 & z + \weps_{-\down} &                  &                    \\
                 &                    & z - \weps_{-\up} &                    \\
                 &                    &                  & z + \weps_{+\down}
\end{pmatrix}
\edq
and 
\beq
\begin{split}
\wsig_0^r &= \sum_{\alpha,k} \wV \hsm_3 \hatg_{\alpha k}^{r} \hsm_3 \wV 
= -\wGamma 
\begin{pmatrix} 
\beta_d & \beta_o \cos(\phi/2) & & \\ 
\beta_o \cos(\phi/2) & \beta_d & & \\
 & & \beta_d & \beta_o \cos(\phi/2) \\ 
 & & \beta_o \cos(\phi/2) & \beta_d
\end{pmatrix}
\end{split}
\edq
The lesser Green's function is given by
\beq
\dbraket{\hatf,\hatf^{\dag}}^< = \dbraket{\hatf,\hatf^{\dag}}^r \wsig_0^< \dbraket{\hatf,\hatf^{\dag}}^a 
= -f(\omega) \left(\dbraket{\hatf,\hatf^{\dag}}^r - \dbraket{\hatf,\hatf^{\dag}}^a\right)
\edq
Eqs.~\eqref{eq:sbse} can be further simplified as
\bes
\begin{align}
\frac{1}{N}  
\int \frac{d\omega}{2\pi i} \tr \left[\wg_f^{r,-1} \dbraket{\hatf,\hatf^{\dag}}_{\omega}^<\right] + \Lambda \frac{\wGamma}{\Gamma} &= 0 \\
\frac{1}{N} \int \frac{d\omega}{2\pi i} \tr \left[\dbraket{\hatf,\hatf^{\dag}}_{\omega}^< \hsm_3 \right] + \frac{\wGamma}{\Gamma} &= \frac{1}{N}.
\end{align}
\eds
Using these equations we obtain analytical expressions for the renormalized parameters $\weps$ and $\wGamma$, from which we can extract the Kondo temperature and the position of the Kondo resonance. The slave boson mean field hamiltonian in Eq. (\ref{eq:mfham}) is quadratic such that we can use the techniques in the previous sections to obtain the Andreev bound states and the Jospehson current. In what follows, we discuss these quantities in different regimes.

First, let us consider the deep Kondo regime in the absence of the spin-orbit coupling $\Delta_{SO} = 0$. Then, the effective level is given by  $\weps = T_{K,SU(4)}$.
Using this fact, the Andreev bound states can be written as
\beq
E_{b}/\Delta = \pm\sqrt{\frac{\weps^2 + \wGamma^2\cos^2(\phi/2)}{\weps^2 + \wGamma^2}}
= \pm \sqrt{\frac{T_{K,SU(4)}^2 + T_{K,SU(4)}^2\cos^2(\phi/2)}{T_{K,SU(4)}^2 + T_{K,SU(4)}^2}}
= \pm \sqrt{\frac{1}{2}\left(1+\cos^2(\phi/2)\right)}
\edq
Employing Eq.~\eqref{eq:Idisab} we find
\beq
I_{dis} = \frac{e\Delta}{2\sqrt{2}\hbar} \sum_{+/-} \frac{\sin(\phi)}{\sqrt{1+\cos^2(\phi/2)}}
= \frac{e\Delta}{\sqrt{2}\hbar} \frac{\sin(\phi)}{\sqrt{1+\cos^2(\phi/2)}}
\edq 
On the contrary, for SU(2) model the effective level is $\weps = 0$
Then,
\beq
E_b/\Delta = \pm \cos(\phi/2)
\edq
which implies
\beq
I_{dis} = \frac{e\Delta}{\hbar} \sin(\phi/2)
\edq

Second, we consider what happens in the presence of the spin-orbit coupling.
In this case, for $T_{K,SU(4)} \gg \Delta_{SO}$, the effective levels become $\weps_{+\up} = \weps_{-\down} = T_{K,SU(4)} + \Delta_{SO}/2$, $\weps_{-\up} = \weps_{+\down} = T_{K,SU(4)} - \Delta_{SO}/2$, such that
\bes
\begin{align}
E_{b+}/\Delta &= \pm\sqrt{\frac{\weps_{+\up}^2 + \wGamma^2\cos^2(\phi/2)}{\weps_{+\up}^2 + \wGamma^2}}
= \pm \sqrt{\frac{\left(1+\frac{\Delta_{SO}}{2T_{K,SU(4)}}\right)^2 + \cos^2(\phi/2)}{\left(1+\frac{\Delta_{SO}}{2T_{K,SU(4)}}\right)^2 + 1}}
\\
E_{b-}/\Delta &= \pm\sqrt{\frac{\weps_{-\up}^2 + \wGamma^2\cos^2(\phi/2)}{\weps_{-\up}^2 + \wGamma^2}}
= \pm \sqrt{\frac{\left(1-\frac{\Delta_{SO}}{2T_{K,SU(4)}}\right)^2 + \cos^2(\phi/2)}{\left(1-\frac{\Delta_{SO}}{2T_{K,SU(4)}}\right)^2 + 1}}
\end{align}
\eds
which yield
\beq
I_{dis} = \frac{e\Delta}{2\hbar} \sum_{\eta=\pm}\frac{\sin(\phi)}{\sqrt{\left[\left(1+\eta\frac{\Delta_{SO}}{2T_{K,SU(4)}}\right)^2 + 1\right]\left[\left(1+\eta\frac{\Delta_{SO}}{2T_{K,SU(4)}}\right)^2 + \cos^2(\phi/2)\right]}}
\edq
Next, we study the opposite limit $T_{K,SU(4)} \ll \Delta_{SO}$. In this case, only the lower level participates in the Kondo physics. Thus,
\beq
\weps_{-+} = \weps_{+\down} = 0 \qquad \text{for $T_{K,SU(2)} \gg \Delta$}
\edq
In this case, the Andreev bound states become $E_{b-}/\Delta = \pm \cos(\phi/2)$ such that
\beq
I_{dis} = \frac{e\Delta}{\hbar} \sin(\phi/2),
\edq
\end{widetext}
with $|\phi|<\pi$ \cite{Beenakker:91}.  
%\bibliography{mybib}

\begin{thebibliography}{25}%
\makeatletter
\providecommand \@ifxundefined [1]{%
 \@ifx{#1\undefined}
}%
\providecommand \@ifnum [1]{%
 \ifnum #1\expandafter \@firstoftwo
 \else \expandafter \@secondoftwo
 \fi
}%
\providecommand \@ifx [1]{%
 \ifx #1\expandafter \@firstoftwo
 \else \expandafter \@secondoftwo
 \fi
}%
\providecommand \natexlab [1]{#1}%
\providecommand \enquote  [1]{``#1''}%
\providecommand \bibnamefont  [1]{#1}%
\providecommand \bibfnamefont [1]{#1}%
\providecommand \citenamefont [1]{#1}%
\providecommand \href@noop [0]{\@secondoftwo}%
\providecommand \href [0]{\begingroup \@sanitize@url \@href}%
\providecommand \@href[1]{\@@startlink{#1}\@@href}%
\providecommand \@@href[1]{\endgroup#1\@@endlink}%
\providecommand \@sanitize@url [0]{\catcode `\\12\catcode `\$12\catcode
  `\&12\catcode `\#12\catcode `\^12\catcode `\_12\catcode `\%12\relax}%
\providecommand \@@startlink[1]{}%
\providecommand \@@endlink[0]{}%
\providecommand \url  [0]{\begingroup\@sanitize@url \@url }%
\providecommand \@url [1]{\endgroup\@href {#1}{\urlprefix }}%
\providecommand \urlprefix  [0]{URL }%
\providecommand \Eprint [0]{\href }%
\@ifxundefined \urlstyle {%
  \providecommand \doi  [0]{\begingroup \@sanitize@url \@doi}%
  \providecommand \@doi [1]{\endgroup \@@startlink {\doibase
  #1}doi:\discretionary {}{}{}#1\@@endlink }%
}{%
  \providecommand \doi  [0]{doi:\discretionary{}{}{}\begingroup
  \urlstyle{rm}\Url }%
}%
\providecommand \doibase [0]{http://dx.doi.org/}%
\providecommand \Doi [0]{\begingroup \@sanitize@url \@Doi }%
\providecommand \@Doi  [1]{\endgroup\@@startlink{\doibase#1}\@@Doi}%
\providecommand \@@Doi [1]{#1\@@endlink}%
\providecommand \selectlanguage [0]{\@gobble}%
\providecommand \bibinfo  [0]{\@secondoftwo}%
\providecommand \bibfield  [0]{\@secondoftwo}%
\providecommand \translation [1]{[#1]}%
\providecommand \BibitemOpen [0]{}%
\providecommand \bibitemStop [0]{}%
\providecommand \bibitemNoStop [0]{.\EOS\space}%
\providecommand \EOS [0]{\spacefactor3000\relax}%
\providecommand \BibitemShut  [1]{\csname bibitem#1\endcsname}%
%</preamble>
\bibitem [{\citenamefont {Kuemmeth}\ \emph {et~al.}(2008)\citenamefont
  {Kuemmeth}, \citenamefont {Ilani}, \citenamefont {Ralph},\ and\ \citenamefont
  {McEuen}}]{Kuemmeth:08}%
  \BibitemOpen
  \bibfield  {author} {\bibinfo {author} {\bibfnamefont {F.}~\bibnamefont
  {Kuemmeth}}, \bibinfo {author} {\bibfnamefont {S.}~\bibnamefont {Ilani}},
  \bibinfo {author} {\bibfnamefont {D.~C.}\ \bibnamefont {Ralph}}, \ and\
  \bibinfo {author} {\bibfnamefont {P.~L.}\ \bibnamefont {McEuen}},\
  }\href@noop {} {\bibfield  {journal} {\bibinfo  {journal} {Nature},\ }\textbf
  {\bibinfo {volume} {452}},\ \bibinfo {pages} {448} (\bibinfo {year}
  {2008})}\BibitemShut {NoStop}%
\bibitem [{\citenamefont {Kasumov}\ \emph {et~al.}(1999)\citenamefont
  {Kasumov}, \citenamefont {Deblock}, \citenamefont {Kociak}, \citenamefont
  {Reulet}, \citenamefont {Bouchiat}, \citenamefont {Khodos}, \citenamefont
  {Gorbatov}, \citenamefont {Volkov}, \citenamefont {Journet},\ and\
  \citenamefont {Burghard}}]{Kasumov:99}%
  \BibitemOpen
  \bibfield  {author} {\bibinfo {author} {\bibfnamefont {A.~Y.}\ \bibnamefont
  {Kasumov}}, \bibinfo {author} {\bibfnamefont {R.}~\bibnamefont {Deblock}},
  \bibinfo {author} {\bibfnamefont {M.}~\bibnamefont {Kociak}}, \bibinfo
  {author} {\bibfnamefont {B.}~\bibnamefont {Reulet}}, \bibinfo {author}
  {\bibfnamefont {H.}~\bibnamefont {Bouchiat}}, \bibinfo {author}
  {\bibfnamefont {I.~I.}\ \bibnamefont {Khodos}}, \bibinfo {author}
  {\bibfnamefont {Y.~B.}\ \bibnamefont {Gorbatov}}, \bibinfo {author}
  {\bibfnamefont {V.~T.}\ \bibnamefont {Volkov}}, \bibinfo {author}
  {\bibfnamefont {C.}~\bibnamefont {Journet}}, \ and\ \bibinfo {author}
  {\bibfnamefont {M.}~\bibnamefont {Burghard}},\ }\href@noop {} {\bibfield
  {journal} {\bibinfo  {journal} {Science},\ }\textbf {\bibinfo {volume}
  {284}},\ \bibinfo {pages} {1508} (\bibinfo {year} {1999})}\BibitemShut
  {NoStop}%
\bibitem [{\citenamefont {Morpurgo}\ \emph {et~al.}(1999)\citenamefont
  {Morpurgo}, \citenamefont {Kong}, \citenamefont {Marcus},\ and\ \citenamefont
  {Dai}}]{Morpurgo:99}%
  \BibitemOpen
  \bibfield  {author} {\bibinfo {author} {\bibfnamefont {A.~F.}\ \bibnamefont
  {Morpurgo}}, \bibinfo {author} {\bibfnamefont {J.}~\bibnamefont {Kong}},
  \bibinfo {author} {\bibfnamefont {C.~M.}\ \bibnamefont {Marcus}}, \ and\
  \bibinfo {author} {\bibfnamefont {H.}~\bibnamefont {Dai}},\ }\href@noop {}
  {\bibfield  {journal} {\bibinfo  {journal} {Science},\ }\textbf {\bibinfo
  {volume} {286}},\ \bibinfo {pages} {263} (\bibinfo {year}
  {1999})}\BibitemShut {NoStop}%
\bibitem [{\citenamefont {Jarillo-Herrero}\ \emph {et~al.}(2006)\citenamefont
  {Jarillo-Herrero}, \citenamefont {Van~Dam},\ and\ \citenamefont
  {Kouwenhoven}}]{Jarillo-Herrero:06}%
  \BibitemOpen
  \bibfield  {author} {\bibinfo {author} {\bibfnamefont {P.}~\bibnamefont
  {Jarillo-Herrero}}, \bibinfo {author} {\bibfnamefont {J.~A.}\ \bibnamefont
  {Van~Dam}}, \ and\ \bibinfo {author} {\bibfnamefont {L.~P.}\ \bibnamefont
  {Kouwenhoven}},\ }\href@noop {} {\bibfield  {journal} {\bibinfo  {journal}
  {Nature},\ }\textbf {\bibinfo {volume} {439}},\ \bibinfo {pages} {953}
  (\bibinfo {year} {2006})}\BibitemShut {NoStop}%
\bibitem [{\citenamefont {Cleziou}\ \emph {et~al.}(2006)\citenamefont
  {Cleziou}, \citenamefont {Wersnsdorfer}, \citenamefont {Bouchiat},
  \citenamefont {Ondarcuhu},\ and\ \citenamefont {Monthioux}}]{Cleziou:06}%
  \BibitemOpen
  \bibfield  {author} {\bibinfo {author} {\bibfnamefont {J.-P.}\ \bibnamefont
  {Cleziou}}, \bibinfo {author} {\bibfnamefont {W.}~\bibnamefont
  {Wersnsdorfer}}, \bibinfo {author} {\bibfnamefont {V.}~\bibnamefont
  {Bouchiat}}, \bibinfo {author} {\bibfnamefont {T.}~\bibnamefont {Ondarcuhu}},
  \ and\ \bibinfo {author} {\bibfnamefont {M.}~\bibnamefont {Monthioux}},\
  }\href@noop {} {\bibfield  {journal} {\bibinfo  {journal} {Nature
  Nanotech.},\ }\textbf {\bibinfo {volume} {1}},\ \bibinfo {pages} {53}
  (\bibinfo {year} {2006})}\BibitemShut {NoStop}%
\bibitem [{Note1()}]{Note1}%
  \BibitemOpen
  \bibinfo {note} {For a review, see S. De Franceschi, L. P. Kouwenhoven, C.
  Schonenberger and W. Wersndorfer, Nature Nanotech., {\protect \bf 5}, 703
  (2010).}\BibitemShut {Stop}%
\bibitem [{\citenamefont {Liang}\ \emph {et~al.}(2002)\citenamefont {Liang},
  \citenamefont {Bockrath},\ and\ \citenamefont {Park}}]{Liang:02}%
  \BibitemOpen
  \bibfield  {author} {\bibinfo {author} {\bibfnamefont {W.}~\bibnamefont
  {Liang}}, \bibinfo {author} {\bibfnamefont {M.}~\bibnamefont {Bockrath}}, \
  and\ \bibinfo {author} {\bibfnamefont {H.}~\bibnamefont {Park}},\ }\Doi
  {10.1103/PhysRevLett.88.126801} {\bibfield  {journal} {\bibinfo  {journal}
  {Phys. Rev. Lett.},\ }\textbf {\bibinfo {volume} {88}},\ \bibinfo {pages}
  {126801} (\bibinfo {year} {2002})}\BibitemShut {NoStop}%
\bibitem [{\citenamefont {Cobden}\ and\ \citenamefont
  {Nyg\aa{}rd}(2002)}]{Cobden:02}%
  \BibitemOpen
  \bibfield  {author} {\bibinfo {author} {\bibfnamefont {D.}~\bibnamefont
  {Cobden}}\ and\ \bibinfo {author} {\bibfnamefont {J.}~\bibnamefont
  {Nyg\aa{}rd}},\ }\href@noop {} {\bibfield  {journal} {\bibinfo  {journal}
  {Phys. Rev. Lett.},\ }\textbf {\bibinfo {volume} {89}},\ \bibinfo {pages}
  {046803} (\bibinfo {year} {2002})}\BibitemShut {NoStop}%
\bibitem [{\citenamefont {Jarillo-Herrero}\ \emph
  {et~al.}(2005){\natexlab{a}}\citenamefont {Jarillo-Herrero}, \citenamefont
  {Kong}, \citenamefont {van~der Zant}, \citenamefont {Dekker}, \citenamefont
  {Kouwenhoven},\ and\ \citenamefont {De~Franceschi}}]{Jarillo-Herrero:05a}%
  \BibitemOpen
  \bibfield  {author} {\bibinfo {author} {\bibfnamefont {P.}~\bibnamefont
  {Jarillo-Herrero}}, \bibinfo {author} {\bibfnamefont {J.}~\bibnamefont
  {Kong}}, \bibinfo {author} {\bibfnamefont {H.~S.~J.}\ \bibnamefont {van~der
  Zant}}, \bibinfo {author} {\bibfnamefont {C.}~\bibnamefont {Dekker}},
  \bibinfo {author} {\bibfnamefont {L.~P.}\ \bibnamefont {Kouwenhoven}}, \ and\
  \bibinfo {author} {\bibfnamefont {S.}~\bibnamefont {De~Franceschi}},\
  }\href@noop {} {\bibfield  {journal} {\bibinfo  {journal} {Nature},\ }\textbf
  {\bibinfo {volume} {434}} (\bibinfo {year} {2005}{\natexlab{a}})}\BibitemShut
  {NoStop}%
\bibitem [{\citenamefont {Choi}\ \emph {et~al.}(2005)\citenamefont {Choi},
  \citenamefont {L\'opez},\ and\ \citenamefont {Aguado}}]{Choi:05}%
  \BibitemOpen
  \bibfield  {author} {\bibinfo {author} {\bibfnamefont {M.-S.}\ \bibnamefont
  {Choi}}, \bibinfo {author} {\bibfnamefont {R.}~\bibnamefont {L\'opez}}, \
  and\ \bibinfo {author} {\bibfnamefont {R.}~\bibnamefont {Aguado}},\ }\Doi
  {10.1103/PhysRevLett.95.067204} {\bibfield  {journal} {\bibinfo  {journal}
  {Phys. Rev. Lett.},\ }\textbf {\bibinfo {volume} {95}},\ \bibinfo {pages}
  {067204} (\bibinfo {year} {2005})}\BibitemShut {NoStop}%
\bibitem [{\citenamefont {Minot}\ \emph {et~al.}(2004)\citenamefont {Minot},
  \citenamefont {Yaish},\ and\ \citenamefont {McEuen}}]{Minot:04}%
  \BibitemOpen
  \bibfield  {author} {\bibinfo {author} {\bibfnamefont {E.}~\bibnamefont
  {Minot}}, \bibinfo {author} {\bibfnamefont {Y.}~\bibnamefont {Yaish}}, \ and\
  \bibinfo {author} {\bibfnamefont {V.~S. . P.~L.}\ \bibnamefont {McEuen}},\
  }\href@noop {} {\bibfield  {journal} {\bibinfo  {journal} {Nature},\ }\textbf
  {\bibinfo {volume} {428}},\ \bibinfo {pages} {536} (\bibinfo {year}
  {2004})}\BibitemShut {NoStop}%
\bibitem [{\citenamefont {Jarillo-Herrero}\ \emph
  {et~al.}(2005){\natexlab{b}}\citenamefont {Jarillo-Herrero}, \citenamefont
  {Kong}, \citenamefont {van~der Zant}, \citenamefont {Dekker}, \citenamefont
  {Kouwenhoven},\ and\ \citenamefont {De~Franceschi}}]{Jarillo-Herrero:05b}%
  \BibitemOpen
  \bibfield  {author} {\bibinfo {author} {\bibfnamefont {P.}~\bibnamefont
  {Jarillo-Herrero}}, \bibinfo {author} {\bibfnamefont {J.}~\bibnamefont
  {Kong}}, \bibinfo {author} {\bibfnamefont {H.~S.~J.}\ \bibnamefont {van~der
  Zant}}, \bibinfo {author} {\bibfnamefont {C.}~\bibnamefont {Dekker}},
  \bibinfo {author} {\bibfnamefont {L.~P.}\ \bibnamefont {Kouwenhoven}}, \ and\
  \bibinfo {author} {\bibfnamefont {S.}~\bibnamefont {De~Franceschi}},\ }\Doi
  {10.1103/PhysRevLett.94.156802} {\bibfield  {journal} {\bibinfo  {journal}
  {Phys. Rev. Lett.},\ }\textbf {\bibinfo {volume} {94}},\ \bibinfo {pages}
  {156802} (\bibinfo {year} {2005}{\natexlab{b}})}\BibitemShut {NoStop}%
\bibitem [{Note2()}]{Note2}%
  \BibitemOpen
  \bibinfo {note} {Various band-structure calculations have been devoted to
  improve the first calculation in T. Ando, J. Phys. Soc. Jpn. {\protect \bf
  69}, 1757 (2000). See e.g. D. Huertas-hernando {\protect \it et al}, Phys.
  Rev. B, {\protect \bf 74}, 155426 (2006); L. Chico {\protect \it et al},
  Phys. Rev. B, {\protect \bf 79} , 235423 (2009)}\BibitemShut {NoStop}%
\bibitem [{\citenamefont {Jespersen}\ \emph {et~al.}(2011)\citenamefont
  {Jespersen}, \citenamefont {Grove-Rasmussen}, \citenamefont {Paaske},
  \citenamefont {Muraki}, \citenamefont {Fujisawa}, \citenamefont
  {Nyg\aa{}rd},\ and\ \citenamefont {Flensberg}}]{Jespersen:11}%
  \BibitemOpen
  \bibfield  {author} {\bibinfo {author} {\bibfnamefont {T.}~\bibnamefont
  {Jespersen}}, \bibinfo {author} {\bibfnamefont {K.}~\bibnamefont
  {Grove-Rasmussen}}, \bibinfo {author} {\bibfnamefont {J.}~\bibnamefont
  {Paaske}}, \bibinfo {author} {\bibfnamefont {K.}~\bibnamefont {Muraki}},
  \bibinfo {author} {\bibfnamefont {T.}~\bibnamefont {Fujisawa}}, \bibinfo
  {author} {\bibfnamefont {J.}~\bibnamefont {Nyg\aa{}rd}}, \ and\ \bibinfo
  {author} {\bibfnamefont {K.}~\bibnamefont {Flensberg}},\ }\href@noop {}
  {\bibfield  {journal} {\bibinfo  {journal} {Nature Phys.},\ }\textbf
  {\bibinfo {volume} {Advance online publication, January}} (\bibinfo {year}
  {2011})}\BibitemShut {NoStop}%
\bibitem [{\citenamefont {Spivak}\ and\ \citenamefont
  {Kivelson}(1991)}]{Spivak:91}%
  \BibitemOpen
  \bibfield  {author} {\bibinfo {author} {\bibfnamefont {B.~I.}\ \bibnamefont
  {Spivak}}\ and\ \bibinfo {author} {\bibfnamefont {S.~A.}\ \bibnamefont
  {Kivelson}},\ }\href@noop {} {\bibfield  {journal} {\bibinfo  {journal}
  {Phys. Rev. B},\ }\textbf {\bibinfo {volume} {43}},\ \bibinfo {pages} {3740}
  (\bibinfo {year} {1991})}\BibitemShut {NoStop}%
\bibitem [{\citenamefont {Vecino}\ \emph {et~al.}(2003)\citenamefont {Vecino},
  \citenamefont {Martin-Rodero},\ and\ \citenamefont
  {Levy~Yeyati}}]{Vecino:03}%
  \BibitemOpen
  \bibfield  {author} {\bibinfo {author} {\bibfnamefont {E.}~\bibnamefont
  {Vecino}}, \bibinfo {author} {\bibfnamefont {A.}~\bibnamefont
  {Martin-Rodero}}, \ and\ \bibinfo {author} {\bibfnamefont {A.}~\bibnamefont
  {Levy~Yeyati}},\ }\href@noop {} {\bibfield  {journal} {\bibinfo  {journal}
  {Phys. Rev. B},\ }\textbf {\bibinfo {volume} {68}},\ \bibinfo {pages}
  {035105} (\bibinfo {year} {2003})}\BibitemShut {NoStop}%
\bibitem [{\citenamefont {Pillet}\ \emph {et~al.}(2010)\citenamefont {Pillet},
  \citenamefont {Quay}, \citenamefont {Morfin}, \citenamefont {Bena},
  \citenamefont {Levy~Yeyati},\ and\ \citenamefont {Joyez}}]{Pillet:10}%
  \BibitemOpen
  \bibfield  {author} {\bibinfo {author} {\bibfnamefont {J.-D.}\ \bibnamefont
  {Pillet}}, \bibinfo {author} {\bibfnamefont {C.~H.~L.}\ \bibnamefont {Quay}},
  \bibinfo {author} {\bibfnamefont {P.}~\bibnamefont {Morfin}}, \bibinfo
  {author} {\bibfnamefont {C.}~\bibnamefont {Bena}}, \bibinfo {author}
  {\bibfnamefont {A.}~\bibnamefont {Levy~Yeyati}}, \ and\ \bibinfo {author}
  {\bibfnamefont {P.}~\bibnamefont {Joyez}},\ }\href@noop {} {\bibfield
  {journal} {\bibinfo  {journal} {Nature Phys.},\ }\textbf {\bibinfo {volume}
  {6}},\ \bibinfo {pages} {965} (\bibinfo {year} {2010})}\BibitemShut {NoStop}%
\bibitem [{\citenamefont {Novotn\'y}\ \emph {et~al.}(2005)\citenamefont
  {Novotn\'y}, \citenamefont {Rossini},\ and\ \citenamefont
  {Flensberg}}]{Novotn:05}%
  \BibitemOpen
  \bibfield  {author} {\bibinfo {author} {\bibfnamefont {T.~c.~v.}\
  \bibnamefont {Novotn\'y}}, \bibinfo {author} {\bibfnamefont {A.}~\bibnamefont
  {Rossini}}, \ and\ \bibinfo {author} {\bibfnamefont {K.}~\bibnamefont
  {Flensberg}},\ }\Doi {10.1103/PhysRevB.72.224502} {\bibfield  {journal}
  {\bibinfo  {journal} {Phys. Rev. B},\ }\textbf {\bibinfo {volume} {72}},\
  \bibinfo {pages} {224502} (\bibinfo {year} {2005})}\BibitemShut {NoStop}%
\bibitem [{\citenamefont {Beenakker}\ and\ \citenamefont {van
  Houten}(1991)}]{Beenakker:91}%
  \BibitemOpen
  \bibfield  {author} {\bibinfo {author} {\bibfnamefont {C.~W.~J.}\
  \bibnamefont {Beenakker}}\ and\ \bibinfo {author} {\bibfnamefont
  {H.}~\bibnamefont {van Houten}},\ }\Doi {10.1103/PhysRevLett.66.3056}
  {\bibfield  {journal} {\bibinfo  {journal} {Phys. Rev. Lett.},\ }\textbf
  {\bibinfo {volume} {66}},\ \bibinfo {pages} {3056} (\bibinfo {year}
  {1991})}\BibitemShut {NoStop}%
\bibitem [{\citenamefont {Zazunov}\ \emph {et~al.}(2010)\citenamefont
  {Zazunov}, \citenamefont {Yeyati},\ and\ \citenamefont
  {Egger}}]{Zazunov:010}%
  \BibitemOpen
  \bibfield  {author} {\bibinfo {author} {\bibfnamefont {A.}~\bibnamefont
  {Zazunov}}, \bibinfo {author} {\bibfnamefont {A.~L.}\ \bibnamefont {Yeyati}},
  \ and\ \bibinfo {author} {\bibfnamefont {R.}~\bibnamefont {Egger}},\ }\Doi
  {10.1103/PhysRevB.81.012502} {\bibfield  {journal} {\bibinfo  {journal}
  {Phys. Rev. B},\ }\textbf {\bibinfo {volume} {81}},\ \bibinfo {pages}
  {012502} (\bibinfo {year} {2010})}\BibitemShut {NoStop}%
\bibitem [{\citenamefont {Bergeret}\ \emph {et~al.}(2010)\citenamefont
  {Bergeret}, \citenamefont {Virtanen}, \citenamefont {Heikkil\"a},\ and\
  \citenamefont {Cuevas}}]{Bergeret:10}%
  \BibitemOpen
  \bibfield  {author} {\bibinfo {author} {\bibfnamefont {F.~S.}\ \bibnamefont
  {Bergeret}}, \bibinfo {author} {\bibfnamefont {P.}~\bibnamefont {Virtanen}},
  \bibinfo {author} {\bibfnamefont {T.~T.}\ \bibnamefont {Heikkil\"a}}, \ and\
  \bibinfo {author} {\bibfnamefont {J.~C.}\ \bibnamefont {Cuevas}},\
  }\href@noop {} {\bibfield  {journal} {\bibinfo  {journal} {Phys. Rev.
  Lett.},\ }\textbf {\bibinfo {volume} {105}},\ \bibinfo {pages} {117001}
  (\bibinfo {year} {2010})}\BibitemShut {NoStop}%
\bibitem [{\citenamefont {Izumida}\ \emph {et~al.}(2009)\citenamefont
  {Izumida}, \citenamefont {Sato},\ and\ \citenamefont {Saito}}]{Izumida:09}%
  \BibitemOpen
  \bibfield  {author} {\bibinfo {author} {\bibfnamefont {W.}~\bibnamefont
  {Izumida}}, \bibinfo {author} {\bibfnamefont {K.}~\bibnamefont {Sato}}, \
  and\ \bibinfo {author} {\bibfnamefont {R.}~\bibnamefont {Saito}},\
  }\href@noop {} {\bibfield  {journal} {\bibinfo  {journal} {J. Phys. Soc.
  Jpn.},\ }\textbf {\bibinfo {volume} {78}},\ \bibinfo {pages} {074707}
  (\bibinfo {year} {2009})}\BibitemShut {NoStop}%
\bibitem [{\citenamefont {Jeong}\ and\ \citenamefont {Lee}(2009)}]{Jeong:09}%
  \BibitemOpen
  \bibfield  {author} {\bibinfo {author} {\bibfnamefont {J.}~\bibnamefont
  {Jeong}}\ and\ \bibinfo {author} {\bibfnamefont {H.}~\bibnamefont {Lee}},\
  }\href@noop {} {\bibfield  {journal} {\bibinfo  {journal} {Phys. Rev. B},\
  }\textbf {\bibinfo {volume} {80}},\ \bibinfo {pages} {075409} (\bibinfo
  {year} {2009})}\BibitemShut {NoStop}%
\bibitem [{\citenamefont {Bulaev}\ \emph {et~al.}(2008)\citenamefont {Bulaev},
  \citenamefont {Trauzettel},\ and\ \citenamefont {Loss}}]{Bulaev:08}%
  \BibitemOpen
  \bibfield  {author} {\bibinfo {author} {\bibfnamefont {D.~V.}\ \bibnamefont
  {Bulaev}}, \bibinfo {author} {\bibfnamefont {B.}~\bibnamefont {Trauzettel}},
  \ and\ \bibinfo {author} {\bibfnamefont {D.}~\bibnamefont {Loss}},\ }\Doi
  {10.1103/PhysRevB.77.235301} {\bibfield  {journal} {\bibinfo  {journal}
  {Phys. Rev. B},\ }\textbf {\bibinfo {volume} {77}},\ \bibinfo {pages}
  {235301} (\bibinfo {year} {2008})}\BibitemShut {NoStop}%
\bibitem [{\citenamefont {{Weiss}}\ \emph {et~al.}(2010)\citenamefont
  {{Weiss}}, \citenamefont {{Rashba}}, \citenamefont {{Kuemmeth}},
  \citenamefont {{Churchill}},\ and\ \citenamefont {{Flensberg}}}]{Weiss:010}%
  \BibitemOpen
  \bibfield  {author} {\bibinfo {author} {\bibfnamefont {S.}~\bibnamefont
  {{Weiss}}}, \bibinfo {author} {\bibfnamefont {E.~I.}\ \bibnamefont
  {{Rashba}}}, \bibinfo {author} {\bibfnamefont {F.}~\bibnamefont
  {{Kuemmeth}}}, \bibinfo {author} {\bibfnamefont {H.~O.~H.}\ \bibnamefont
  {{Churchill}}}, \ and\ \bibinfo {author} {\bibfnamefont {K.}~\bibnamefont
  {{Flensberg}}},\ }\Doi {10.1103/PhysRevB.82.165427} {\bibfield  {journal}
  {\bibinfo  {journal} {\prb},\ }\textbf {\bibinfo {volume} {82}},\ \bibinfo
  {pages} {165427} (\bibinfo {year} {2010})}\BibitemShut {NoStop}%
\end{thebibliography}

%merlin.mbs 2010-03-15 4.21a (PWD, AO, DPC)
\end{document}